\documentclass[pdflatex,11pt,en]{aghdpl}
\pdfoutput=1 

\usepackage[english]{babel}
\usepackage{enumerate}
\usepackage{listings}
\usepackage{framed}
\usepackage{multirow}
\usepackage{array}
\usepackage{booktabs}
\usepackage{tabularx}
\usepackage[table]{xcolor}
\usepackage{amsthm}
\usepackage{wrapfig}
\usepackage{rotating}
\usepackage{fancyvrb}
\usepackage{pdflscape}
\usepackage{afterpage}
\usepackage[labelsep=quad]{caption} 
\usepackage{bookmark}
\usepackage{enumitem} 

\addto\captionsenglish{

}

\newtheorem{definition}{Definition}
\bookmarksetup{open,numbered}

\author{Jarosław Waliszko}
\email{jaroslaw.waliszko@gmail.com}
\shortauthor{J. Waliszko}
\title{Knowledge representation and processing methods in Semantic Web} 
\shorttitle{Knowledge representation and processing methods in Semantic Web}
\thesistype{Master of Science Thesis}
\supervisor{Antoni Ligęza, Professor}
\date{2010}
\faculty{Faculty of Electrical Engineering, Automatics, Computer Science and Electronics}
\department{Department of Automatics}
\degreeprogramme{Applied Computer Science}
\acknowledgements{I gratefully acknowledge the help and support of my advisor, Professor Antoni Ligęza. Without his assistance the work embodied in this thesis would not have been possible. Thanks are also due to my Parents for support and understanding.}
\thesiskeywords{semantic web, knowledge representation, description logic, ontology, ontology driven application, web system, database, reasoning, traffic dangers}

\begin{document}

\titlepages
\tableofcontents

\clearpage
\chapter{Introduction}
\label{cha:introduction}

\section{Goals}
\label{sec:goals}

The goal of this thesis is to take a closer look at progress of knowledge engineering in the field of Semantic Web. Along with theory of Knowledge Representation (KR) \cite{HLP08} and knowledge processing methods such as Description Logic (DL) \cite{BCM03}, reasoning mechanisms and ontology modeling languages (OWL \cite{W3COWL}, RDF \cite{RDFPrimer}, RDFS \cite{RDFSchema}), the thesis shows the practical usage of the mentioned approaches in building systems driven by ontologies. 

A working prototype of ontology-driven application, written in Java, has been developed within the scope of this thesis. The system main assumption is an attempt to integrate database and ontology approach, for storing and inferring desired information about domain of traffic dangers. For the needs of the system, domain model of traffic danger concept has been also designed. The ontology has been built using Protégé \cite{ProtegeHome} editor integrated with description logic reasoner. 

In presented solution, ontology domain description is supported by data stored in database. Location details of traffic conditions are stored in database, while clean abstract of traffic danger domain is described by ontology. Such integration results in dynamic deduction possibility of specific facts, using DL reasoning services, instead of using only static relations defined in database. Ontology-based approach gives the meaning to the information.

\newpage

\section{Content}
\label{sec:content}

Chapter \ref{cha:introduction} briefly outlines the goals and contents of the thesis.

\bigskip

\noindent Chapter \ref{cha:theoreticalBackground} introduces the concept of Semantic Web, Knowledge Representation languages, Description Logics and reasoning mechanisms. Future goals of Semantic Web applications is shown along with explanation why it is worth to develop ontologies.

\bigskip

\noindent Chapter \ref{cha:trafficDangerOntology} explains the traffic danger concept. Ontology development process based on Protégé tool from Stanford Center for Biomedical Informatics Research is shown and described. Protégé and knowledge engineering approach used for ontology development is introduced.

\bigskip

\noindent Chapter \ref{cha:trafficDangerWebSystem} shows ontology-based reasoning system built as a proposal of knowledge processing application. The system works as a web application and integrates database approach with ontology-based approach for combining knowledge. It can dynamically infer answers for user defined questions in real time. Particular parts of the application are explained. Model-driven architecture along with agile methodology is introduced. Technologies used in building process are also briefly outlined.

\bigskip

\noindent Chapter \ref{cha:conclusion} shortly summarizes the work.

\chapter{Theoretical background}
\label{cha:theoreticalBackground}

\section{Semantic Web}
\label{sec:semanticWeb}

\subsection{Background}
\label{sub:semWebBackground}

Today's World Wide Web has achieved a great success. WWW is widely used all around the world. Simple foundations, on which it is based, has given its strength and global community. Almost every user is able to create and publish hypertext resources in a simple manner. Basically \textit{"web content consists mainly of distributed hypertext, and is accessed via a combination of keyword-based search and link navigation"} \cite{HorSSF}.

Notably, upon explosion of information amount accessible through the web, current web has encountered drawbacks. The huge amount of content required to be processed in order to find desired facts, has \textit{"highlighted some serious shortcomings in the hypertext paradigm"} \cite{HorSSF}:
\begin{itemize}
    \setlength{\itemsep}{0cm}
    \setlength{\parskip}{0cm}

    \item difficulties in finding correct information through simple searching and browsing mechanisms,
    \item problems of finding facts, which has common correspondents,
    \item irrelevant results after providing of more complex queries in the browsing engines,
    \item lack of semantics,
    \item lack of deduction facilities.
\end{itemize}

\noindent People are facing inconveniences in web content browsing. It is obvious, that content processing tasks are much more complicated when it comes to implement them into software agents. It is because technologies used for hypertext documents creation are designed with the intention of describing and presenting information in the human way. The lack of formal, logic-based information structure is the big issue for machines, which require mathematically-based way for knowledge processing. Here are the challenges, current web is going to face.

\newpage

\subsection{Evolution}
\label{sub:semWebEvolution}

Today's World Wide Web gives new opportunities for discipline called Knowledge Representation (KR). Web resources needs to be better adopted for automatic processing by computer agents. That is possible by more systematic and formal description mechanisms used for defining contents of the pages. What is more, Uniform Resource Identifiers (URIs), which are the naming scheme for web resources, allows KR systems to improve linking between semantic documents by avoiding the ambiguities of natural language. That is why new possibilities and challenges are opened for knowledge representation assumptions \cite{HLP08}.

One of the most powerful feature of the World Wide Web is the possibility of interconnections between knowledge sources. Interconnections are provided by linking mechanisms on pages. In other words, instead of copying information from external sources created by other people to our page, we just provide links to such sources.

Semantic Web can be treated as an extension to current web and the successor of Web 2.0. In recent Web 2.0 architecture, there is a lot of unstructured data. Information is represented in various ways, incompatible witch each other. What Semantic Web offers is defined meaning of information, machine-readable description of contents, simplification of information exchange, classification and inference mechanisms, data processing and integration in machine-automated manner. 

\medskip

\begin{figure}[htp]
\centering
\includegraphics[scale=0.65]{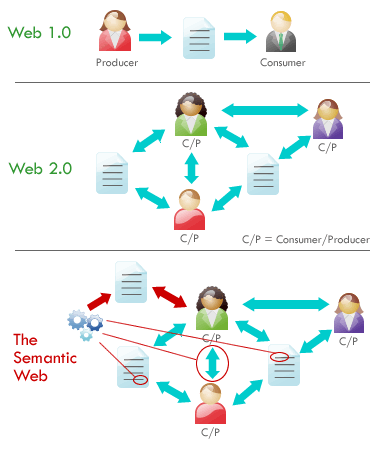}
\caption{Web evolution from old Web 1.0 up to Semantic Web \cite{PWGB}}
\label{fig:webEvolution}
\end{figure}

\newpage

\subsection{Outline}
\label{sub:semWebOutline}

\noindent \textit{"The Semantic Web is an extension of the current Web in which information is given well-defined meaning, better enabling computers and people to work in cooperation."} 
\begin{flushright} Tim Berners-Lee \cite{BHL01} \end{flushright}

\noindent \textit{"The goal of Semantic Web research is to transform the Web from a linked document repository into a distributed knowledge base and application platform, thus allowing the vast range of available information and services to be more effectively exploited."} 
\begin{flushright} Ian Horrocks \cite{HorSSF} \end{flushright}

\noindent The Semantic Web is not about links between web pages, but about relationships between things (e.g. A is a part of B and Y is a member of Z) and about properties of things (e.g. size, weight, age and price) \cite{SemWebTutorial}.

\bigskip

\noindent \textit{"If HTML and the Web made all the online documents look like one huge book, RDF, schema, and inference languages will make all the data in the world look like one huge database."}
\begin{flushright} Tim Berners-Lee \cite{BLF99} \end{flushright}

\noindent As it was said, the Semantic Web is a web that is able to provide information about things in the way computers can understand. The human-understandable sentences which describe things, are going to be intelligently processed in the Semantic Web. 

Statements appropriate for dedicated language are built using syntax rules, which are defined by this language. In Semantic Web field statements are built using rules provided by specific KR languages and such statements are semantic aware. Semantic analysis of such statements allows to convert them to knowledge base queries. To search and access the Semantic Web content, tools called \textit{Semantic Web Agents} or \textit{Semantic Web Services} are required. It is because searching process in Semantic Web will be much different than today's searching in free text - complicated math algorithms are going to find answers for us.

\bigskip

\noindent \textit{"The Semantic Web provides a common framework that allows data to be shared and reused across application, enterprise, and community boundaries."}
\begin{flushright} World Wide Web Consortium \cite{SemWebActivity} \end{flushright}

\newpage

\begin{figure}[htp]
\centering
\includegraphics[scale=0.6]{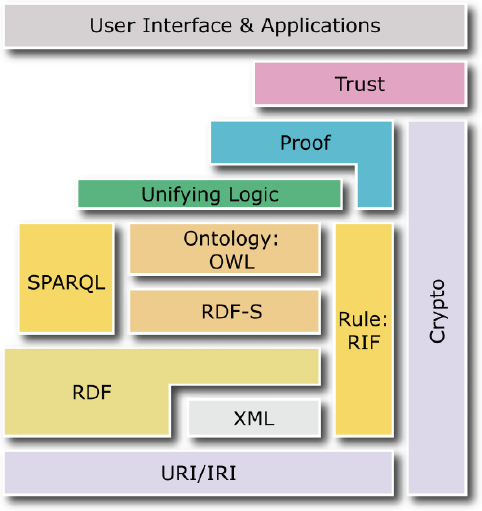}
\caption{Semantic Web Stack \cite{GJNSemWeb}}
\label{fig:semWebStack}
\end{figure}

\noindent The Semantic Web layered architecture (Figure \ref{fig:semWebStack}) covers such technological aspects like application interfaces, security, rules and logic (SWRL, RIF), ontologies (RDFS, OWL), metadata (RDF), serialization (XML) and resource identification (URI).

\medskip

\subsubsection{KR languages and tools}
\label{sss:krLanguagesAndTools}

The KR languages: RDF, RDFS and OWL explained in the next sections, are \textit{"without question the most widely used KR languages in history"} \cite{HLP08}. They were created to capture and describe knowledge gathered by the web resources, and help applications to use these resources in an intelligent way. These languages are shaping the new standards in web development, allowing for transformation to the new standards. The Web 3.0 is slowly coming to light.

Along with the mentioned languages, development of new tools cooperating with them has appeared. Tools are designed for creating and maintaining taxonomies, and are supported by DL reasoning services (see Section \ref{sub:owlAndReasoners}). It gives the wide group of Semantic Web community new possibilities and motivation for creation ontology-based applications and experimenting with that technology. New challenges have appeared such as large scale OWL-based applications development.

\newpage

\section{Towards ontology development}
\label{sec:towardsOntologyDevelopment}

\subsection{Introduction}
\label{sub:towardsOntologiesDevelopmentIntroduction}

\textit{"In recent years the development of ontologies has been moving from the realm of Artificial-Intelligence laboratories to the desktops of domain experts"} \cite{OntDev101}. Many organizations and disciplines has developed standardized ontologies that domain experts can exchange and reuse in building their own ones. An ontology, \textit{"explicit formal specifications of the terms in the domain and relations among them"} \cite{Gru93}, specifies set of common vocabulary, researchers can share within a domain of knowledge. Ontologies have become increasingly important with the global move towards the Semantic Web. Parallel with Semantic Web evolution, knowledge defined by ontologies can be available on the web to a multitude of machines. 

\subsection{Reasons and advantages}
\label{sub:reasonsAndAdvantages}

The key reasons for developing an ontology \cite{OntDev101} are following:
\begin{enumerate}
    \setlength{\itemsep}{0cm}
    \setlength{\parskip}{0cm}

    \item To share common understanding of the structure of information among people or software agents.
    \item To enable reuse of domain knowledge.
    \item To make domain assumptions explicit.
    \item To separate domain knowledge from the operational knowledge.
    \item To analyze domain knowledge.
\end{enumerate}

Ad 1. Sharing common understanding of knowledge domain is one of the most critical reason for developing ontologies. For example, we can imagine several web sites containing specific technical information. If that information is published under common comprehensive ontology, the software agents will be able to aggregate and process knowledge from them, and infer exhaustive answers for user queries to that domain. 

Ad 2. Coherent domain knowledge sharing, along with reusing of such knowledge can provide the possibility of injecting new aspects in existing set of information. This possibility, will provide complex description of specific knowledge structures to researchers all around the world, where one team of experts can supplement existing structure with new elements, with standardized vocabulary set. Along with automatic processing of knowledge, answering to questions about domain related topics, is going to be coherent and comprehensive.

Ad 3. Explicit domain assumptions provide clear description of domain knowledge, and simplify knowledge extensibility for domain. New researchers have the possibility of view to comprehensive domain related information structure, and while learning about that structure, have easier task in understanding what specific terms mean. Besides, in the world of software applications, explicit assumptions are much better than hard coded assumptions. Hard coded information along with all relations, would be much harder to find, and any changes would be impossible to execute, without programming expertise.

Ad 4. When it comes to separation, between domain knowledge and operational knowledge, we can talk about some kind of loose coupling between them. We can develop abstract set of assumptions in the domain, and next use that metadata for providing specific, context-based information about such domain. For example domain experts can research an ontology which describe traffic dangers aspect, and users can provide specific traffic danger information for that underlying ontology, by defining concrete locations where specific traffic conditions can appear.

Ad 5. Analyzing domain knowledge is an obvious value on its own. It is critical when it comes to extend existing, or providing new branches of specific domain. Analysis should be based on comprehensive, structured source of knowledge, which an ontology is, and should provide an easy way of global view of concept along with description of all related terms used to define it.

\bigskip

\noindent Similar classification provided by The Knowledge Systems, AI Laboratory (KSL) at Stanford University \cite{KSL} is following:
\begin{itemize}
    \setlength{\itemsep}{0cm}
    \setlength{\parskip}{0cm}

    \item to enable a machine to use the knowledge in some application,
    \item to enable multiple machines to share their knowledge,
    \item to help yourself understand some area of knowledge better,
    \item to help other people understand some area of knowledge,
    \item to help people reach a consensus in their understanding of some area of knowledge.
\end{itemize}

\noindent These particular points, in a bit different form, were described earlier. Ontology development aims, described by different sources, are convergent. Ontologies development seems to be crucial in the near future, because the size of human knowledge is going to be too wide for processing in the meaning of standard way.

\newpage

\section{XML}
\label{sec:xml}

\subsection{Introduction}
\label{sub:xmlIntroduction}

The starting point of introduction to world of KR languages, should be brief description of XML language. It is because XML lies on the bottom of Semantic Web layered stack and provides the serialization for mentioned KR languages.

XML is widely used technology so information stored in variety of papers, e.g. \cite{W3SchoolsXML, W3CXML}, can be easily found to help understand, what exactly is XML.

\bigskip

\noindent XML facts:
\begin{itemize}
    \setlength{\itemsep}{0cm}
    \setlength{\parskip}{0cm}

    \item XML stands for EXtensible Markup Language,
    \item XML is a markup language much like HTML,
    \item XML allows to represent tree-like structures,
    \item XML was designed to carry data, not to display data,
    \item XML tags are not predefined - custom tags have to be defined,
    \item XML is designed to be self-descriptive,
    \item XML is platform independent,
    \item XML files are stored as plain text files,
    \item XML emphasizes simplicity, generality, and usability over internet,
    \item XML is a W3C Recommendation.
\end{itemize}

\noindent XML is used in many various aspects of coherent data transmission through the web. In the real world incompatible data formats coexist in computer systems, and are completely incomprehensible to each other. XML provides common, interchangeable way of storing and sharing data. It is a software and hardware independent tool for carrying information. It is also used as the serialization mechanism for various, newly created internet languages like: XHTML, WSDL (describing Web Services), RSS (used for news feeds), RDF and OWL (describing resources and ontologies), SMIL (describing multimedia for the web), etc.

\newpage

\noindent Below, there is a sample XML document:

{\tt \small
\begin{verbatim}
<?xml version="1.0" encoding="UTF-8"?>
<message status="critical" releaseDate="13th April, 1970">
   <to>Manned Spacecraft Center (building 30), Houston, Texas</to>
   <from>Commander James A. Lovell, Apollo 13 crew</from>
   <heading>Technical Fault</heading>
   <body>
      Houston, we've had a problem. 
      We've had a main B bus undervolt.
   </body>
</message>
\end{verbatim}
}

\noindent First line of the file indicates, that it is version 1.0 of XML standard, and information is encoded with UTF-8 character encoding. File content declares a message sent on the 13th of April, 1970, from Apollo 13 spacecraft to mission control center, about technical issue connected with oxygen tank explosion. 

This document is quite self-descriptive. It contains sender and receiver information, heading and the message body. Notably, that piece of code does not provide any functionality, besides carrying and structuralizing the data. It is just information wrapped in tags. There is still a necessity for writing software which will be able to process and respond, or display that information in more friendly way.

\subsection{XML syntax}
\label{sub:xmlSyntax}

\subsubsection{Elements}
\label{sss:xmlElements}

The tags used inside XML documents, are not defined in any XML standard. XML allows developers to invent custom tags, because as opposed to HTML where tags are the part of the standard, XML has no predefined tags.

XML documents have to contain one root element. Structure of XML documents is the structure of rooted tree. The tree starts at the root and branches to the lowest level of the tree. A rooted tree is a connected, acyclic, undirected graph, in which one of the vertices is distinguished from others. The distinguished vertex is called the root of the tree \cite{CLR90}.

\newpage

{\tt \small
\begin{verbatim}
<root>
   <firstChild>
      <subChild>.....</subChild>
   </firstChild>
   <secondChild>
      <subChild>.....</subChild>
   </secondChild>
</root>
\end{verbatim}
}

\noindent An XML element is everything from (including) the element start tag to (including) the element end tag. Elements can contain other elements, plain text or a mixture of both. Each XML tag have to close previously opened tag, in the correct order. It is illegal to have unclosed tag in XML document. All elements must be properly nested. What is more, XML is case sensitive.

{\tt \small
\begin{verbatim}
<p>incorrect - no closing tag
<p>correct</p>

<p><b>incorrect - improperly nested elements</p></b>
<p><b>correct</b></p>

<P>incorrect - tags are not matched in the manner of case sensitivity</p>
<P>correct</P>
<p>correct</p>
\end{verbatim}
}

\subsubsection{Attributes}
\label{sss:xmlAttributes}

XML elements can have attributes. Attributes provide additional information about elements. Attribute is a simple key-value pair, in the format of \textit{key="value"}. Attribute values have to be quoted.

{\tt \small
\begin{verbatim}
<message status="critical">.....</message>
\end{verbatim}
}

\noindent There are some problems with attributes. In opposite to elements they cannot contain multiple values, as well as tree structures. They are not easily expandable for future changes. 

\setlist{nosep} 
\begin{framed}
\begin{itemize}
    \setlength{\itemsep}{0cm}
    \setlength{\parskip}{0cm}

    \item Attributes should be used for providing information, which is not relevant to the data itself. They are excellent for storing \textit{metadata} (data about data) (such as ID of message, etc.).
    \item For storing \textit{data itself}, elements are much better choice than attributes.
\end{itemize}
\end{framed}
\setlist{} 

\newpage

\subsubsection{Others}
\label{sss:xmlOthers}

XML supports comments. The syntax for writing comments in XML is similar to that of HTML:

{\tt \small
\begin{verbatim}
<!-- This is a comment --> 
\end{verbatim}
}

\noindent Some characters in XML Standard have special purpose. It means, that XML parsers try to interpret such characters in a special way. To insert such a character in the XML document without its meaning, predefined entity references should be used:

\medskip

\begin{table}[htp]
\centering
\begin{tabular}{ |>{\tt}p{2cm}|>{\tt}l|l| }
    \hline
    \&lt;       & <     & less than \\ \hline
    \&gt;       & >     & greater than \\ \hline
    \&amp;      & \&    & ampersand \\ \hline
    \&apos;     & '     & apostrophe \\ \hline
    \&quot;     & "     & quotation mark \\ \hline
\end{tabular}
\caption{Predefined entities in XML 1.0}
\end{table}

\noindent As opposed to HTML, white-space characters in XML document are NOT truncated. When it comes to storing new lines, XML behaves in the same way as Unix/Mac applications, and uses line feed character (LF).

XML documents can follow a set of strict rules, to store information in a predefined format. If they follow such rules, they are \textit{valid}. Such predefined data structure is critical during parsing of XML documents by various applications. It helps applications to properly interpret specific information, because when transporting data, it is essential that both sender and receiver have the same "expectations" about the content. If XML document doesn't follow its schema, it can be rejected by application as invalid. 

\subsection{Structure and content validation}
\label{sub:dtd}

The standard defining syntax rules which should be followed by XML documents is called Document Type Definition (DTD). It describes the structure with a list of legal tags and their attributes. DTD can be embedded within XML file: 

\newpage

{\tt \small
\begin{verbatim}
<?xml version="1.0" encoding="UTF-8"?>
<!DOCTYPE message [
<!ELEMENT message (to,from,heading,body)>
<!ATTLIST message status (spam|normal|critical) "spam">
<!ATTLIST message releaseDate CDATA #REQUIRED>
<!ELEMENT to (#PCDATA)>
<!ELEMENT from (#PCDATA)>
<!ELEMENT heading (#PCDATA)>
<!ELEMENT body (#PCDATA)>
]>
<message status="critical" releaseDate="13th April, 1970">
   <to>Manned Spacecraft Center (building 30), Houston, Texas</to>
   <from>Commander James A. Lovell, Apollo 13 crew</from>
   <heading>Technical Fault</heading>
   <body>
      Houston, we've had a problem. 
      We've had a main B bus undervolt.
   </body>
</message>
\end{verbatim}
}

\noindent or can exist as a separate file, which can be referenced from the XML document, as marked out below:

{\tt \small
\newcommand\codeHighlight[1]{\textcolor[rgb]{1,0,0}{\textbf{#1}}}
\begin{Verbatim}[commandchars=\\\{\}]
<?xml version="1.0" encoding="UTF-8"?>
\textbf{<!DOCTYPE message SYSTEM "\codeHighlight{message.dtd}">}
<message status="critical" releaseDate="13th April, 1970">
   <to>Manned Spacecraft Center (building 30), Houston, Texas</to>
   <from>Commander James A. Lovell, Apollo 13 crew</from>
   <heading>Technical Fault</heading>
   <body>
      Houston, we've had a problem. 
      We've had a main B bus undervolt.
   </body>
</message>
\end{Verbatim}
}

\noindent More powerful XML-based alternative to DTD is called XML Schema. The migration from older DTDs to XML Schema includes following benefits:
\begin{itemize}
    \setlength{\itemsep}{0cm}
    \setlength{\parskip}{0cm}

    \item Schemas support primitive (built-in) data types (e.g. \texttt{xsd:integer}, \texttt{xsd:string}, \texttt{xsd:date}, etc.) and data namespaces.
    \item Schemas are extensible. Schemas gives possibility of defining custom data types, using object-oriented data modeling principles: encapsulation, inheritance and substitution.
    \item It is possible to reuse them in other Schemas, create new data types derived from the standard types and reference multiple schemas within the same document.
    \item Schemas are compatible with other XML technologies (Web Services, XQuery, XSLT, etc.).
\end{itemize}

\noindent Sample schema shown below, can be used as an alternative to DTD introduced earlier, for defining the syntactic correctness of XML document also discussed earlier:

{\tt \small
\begin{verbatim}
<xs:element name="message">
   <xs:complexType>
      <xs:sequence>
         <xs:element name="to" type="xs:string"/>
         <xs:element name="from" type="xs:string"/>
         <xs:element name="heading" type="xs:string"/>
         <xs:element name="body" type="xs:string"/>
      </xs:sequence>
      <xs:attribute name="status" type="xs:string" default="spam"/>
      <xs:attribute name="releaseDate" type="xs:string" use="required"/>
   </xs:complexType>
</xs:element>
\end{verbatim}
}

\section{RDF}
\label{sec:rdf}

\subsection{Introduction}
\label{sub:rdfIntroduction}

If we take a look once again at Semantic Web Stack, we can see that just above XML lies Resource Description Framework (RDF). RDF is a simple assertional language used for describing resources in the World Wide Web. RDF is intended for situations where information is processed in automatic manner by applications, rather than being displayed to users. It is designed to represent information in the form of \textbf{triples}, such as \textit{subject}, \textit{predicate} and \textit{object}. RDF predicates can be treated as attributes of resources. Sample \textbf{subject-predicate-object} (object-attribute-value) triple building block is shown on the directed graph below:

\medskip

\begin{figure}[htp]
\centering
\includegraphics[scale=0.72]{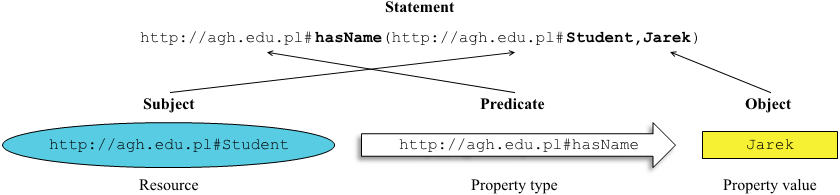}
\caption{Example of RDF statement}
\label{fig:rdfTriple}
\end{figure}

\newpage

\noindent Resources refers to things being described. Resources are identified by URIs. The binary relations between resources are called properties. RDF properties themselves also have URIs, to be more accurate in the identification of the relations between resources. Statements are created as the asserted properties of resources.

\subsection{RDF syntax}
\label{sub:rdfSyntax}

RDF provides XML-based syntax, called RDF\slash XML. A piece of RDF\slash XML code written below corresponds to the graph in Figure \ref{fig:rdfTriple}:

{\tt \small
\begin{verbatim}
1. <?xml version="1.0"?>
2. <rdf:RDF xmlns:rdf="http://www.w3.org/1999/02/22-rdf-syntax-ns#"
3.          xmlns:agh="http://agh.edu.pl#">

4.    <rdf:Description rdf:about="http://agh.edu.pl#Student">
5.       <agh:hasName>"Jarek"</agh:hasName>
6.    </rdf:Description>

7. </rdf:RDF>
\end{verbatim}
}

\noindent Line numbers are added to the example. It is because they are referenced in the explanation part below:
\begin{itemize}
    \setlength{\itemsep}{0cm}
    \setlength{\parskip}{0cm}

    \item \textit{Line 1} contains standard XML declaration \texttt{<?xml version="1.0"?>}, which indicates that the content of that file is XML, and provides XML version which is used inside.	
    \item \textit{Line 2} starts from tag \texttt{rdf:RDF}, which indicates that the following XML content syntax is RDF. The \texttt{xmlns:rdf} defines a namespace identified by the URI \url{http://www.w3.org/1999/02/22-rdf-syntax-ns#}, and tells that all tags prefixed with \texttt{rdf:} are parts of the namespace. That namespace is used for terms from RDF vocabulary.
    \item \textit{Line 3} defines another prefix \texttt{agh:}, which represents namespace \url{http://www.agh.edu.pl#}. URI \url{http://agh.edu.pl#} is used for vocabulary terms defined by organization \texttt{agh.edu.pl}.
    \item \textit{Lines 4-6} provide RDF/XML code, which describes the statement shown at graph \ref{fig:rdfTriple}. \textit{Line 4} begins from tag \texttt{rdf:Description} which indicates the start of description of a resource. Next, using attribute \texttt{rdf:about}, identifies the resource the statement is about (the subject of the statement), by providing its URI \url{http://agh.edu.pl#Student}. \textit{Line 5} declares property element \texttt{agh:hasName} for the subject resource, where both predicate and object of the statement are represented. The value of the property identified by namespace \url{http://agh.edu.pl#hasName} is a plain literal \texttt{Jarek}. \textit{Line 6} closes \texttt{rdf:Description} element.
    \item \textit{Line 7} indicates the end of \texttt{rdf:RDF} element.
\end{itemize}

\newpage

\noindent \textbf{RDF vocabulary} is a defined set of predicates that can be used in an application. The vocabulary defined by the RDF specification is following: \texttt{rdf:type} (predicate indicating that resource is an instance of class), \texttt{rdf:XMLLiteral} (class of typed literals), \texttt{rdf:Property} (class of properties), \texttt{rdf:Alt}, \texttt{rdf:Bag}, \texttt{rdf:Seq} (containers), \texttt{rdf:List} (class of lists), \texttt{rdf:nil} (instance of \texttt{rdf:List}, representing empty list), \texttt{rdf:Statement}, \texttt{rdf:subject}, \texttt{rdf:predicate}, \texttt{rdf:object} (used for \textit{reification}, in which each statement (each triple) is assigned a URI and treated as a resource about which further statements can be made). Vocabulary shown here is used as a backbone for RDF Schema, where that limited vocabulary is extended.

RDF\slash XML is machine processable, just like HTML. By using URIs, it can link pieces of information across the web. Unlike conventional hypertext documents, URIs are used by RDF for any identifiable things, even those which are not directly retrievable on the web (such as the student \texttt{Jarek}). In addition to describing web resources (like pages, images, videos, etc.), RDF can also describes abstract things like people, robots, planets, events, etc. \cite{RDFPrimer}

\section{RDFS}
\label{sec:rdfs}

\subsection{Introduction}
\label{sub:rdfsIntroduction}

RDF vocabulary description language (RDF Schema), built on top of RDF, is a general-purpose language for representing information on the web \cite{RDFSchema}. 

RDF properties may be thought of as attributes of resources. In this sense they can be understand as traditional attribute-value pairs. In addition, RDF properties represent relationships between resources. RDF however, provides no mechanisms for describing these properties, nor does it provide any mechanisms for describing the relationships between these properties and other resources. That is the role of the RDF vocabulary description language (RDFS) \cite{RDFSchema}.

RDF Schema extends RDF vocabulary to allow describing taxonomies of classes and properties. Classes (generalized categories or unary relations) and properties (predicates or binary relations) can be arranged into hierarchies. In addition it extends definitions for some of the elements of RDF, i.e. it sets the domain and range of properties \cite{SemWebIntro}. RDFS also gives simple inferencing possibility of the following forms: inferring class membership and subclass relations through transitive inference in the subclass hierarchy, inferring class membership through occurrence in typed property-positions, and inferring property values and subproperty relations through transitive inference in the subproperty hierarchy. 

\newpage

\subsection{RDFS constructs}
\label{sub:rdfsConstructs}

\noindent Main constructs defined by RDFS are:
\begin{itemize}
    \setlength{\itemsep}{0cm}
    \setlength{\parskip}{0cm}

    \item \textbf{Classes}: \texttt{rdfs:Resource} (because all things described by RDF are resources, it is the class of everything), \texttt{rdfs:Class} (declares a resource as a class of another resource), \texttt{rdfs:Literal} (literal values, such as strings or integers), \texttt{rdfs:Datatype} (class of datatypes), \texttt{rdf:XMLLiteral} (the class of XML literal values) and \texttt{rdf:Property} (class of properties). Example:

{\tt \small
\begin{verbatim}
<rdf:Description rdf:about="http://agh.edu.pl#Student">
   <rdf:type rdf:resource=People/>
</rdf:Description>
\end{verbatim}
}

{\tt \small
\begin{verbatim}
<rdfs:Class rdf:ID="Keyboard">
   <rdfs:subClassOf rdf:resource="Notebook"/>
</rdf:Class>
\end{verbatim}
}

    \item \textbf{Properties}: \texttt{rdfs:domain} (declares the class of the \texttt{subject} in a triple block), \texttt{rdfs:range} (declares the class or datatype of the object in a triple block), \texttt{rdf:type} (property used to state that a resource is an instance of a class), \texttt{rdfs:subClassOf} (allows to declare hierarchies of classes) and \texttt{rdfs:subPropertyOf} (allows to declare hierarchies of properties). Example:

{\tt \small
\begin{verbatim}
<rdf:Property rdf:ID="hasUniqueSkills">
   <rdfs:subPropertyOf rdf:resource="hasFeature"/>
   <rdfs:domain rdf:resource="People"/>
   <rdfs:ranage rdf:resource="AGHStudents"/>
</rdf:Property>
\end{verbatim}
}

    \item \textbf{Utility Properties}: \texttt{rdfs:seeAlso} (relates a resource to another with additional explanation), \texttt{rdfs:isDefinedBy} (relates a resource to its definition), \texttt{rdfs:label} (provides friendly name of resource) and \texttt{rdfs:comment} (provides a human-readable description of a resource).
\end{itemize}

\medskip

\noindent The fact is, that there is lack of any notion of negation or disjunction in RDFS. It provides only a very limited notion of existential quantification. This makes for RDFS language \textit{"very limited expressive power"} \cite{HLP08}.

\newpage

\section{DL}
\label{sec:dl}

\subsection{Introduction}
\label{sub:DLIntroduction}

Description Logics (DLs) are family of knowledge representation (KR) languages. They can be used for representing knowledge of application domains in structured, formal way. The first part of the name \textit{description logics} means, that the knowledge is represented by concept \textit{descriptions}. Elementary descriptions are the expressions built from \textit{atomics concepts} (unary predicates) and \textit{atomic roles} (binary predicates). Complex descriptions are created from elementary descriptions inductively by using \textit{concept constructors}. The second part indicates, that description logics are equipped with a formal \textit{logic}-based semantics, unlike their predecessors: semantic networks and frames. DLs were introduced to KR systems as a reaction to overcome deficiencies derived from the lack of formal semantics. One of the key DLs features is the fact, that they provide reasoning algorithms for the composition of structured concept descriptions. Reasoning allows to infer implicitly represented knowledge from the explicitly provided information stored in the knowledge base \cite{BCM03, Hor97, HLP08}.

Reasoning processes allow for classification of concepts and individuals. Classification of concepts determines the dependencies between them, thus creates hierarchies of concepts. Classification of individuals indicates whether given individual is an instance of certain concept. Such inferential mechanisms are used by many intelligent information processing systems. What is more, classifying and ordering attempts of every part of the world is the way humans try to understand reality.

\subsection{DL Knowledge Representations Systems}
\label{sub:DLKRSs}

\begin{figure}[htp]
\centering
\includegraphics[scale=0.6]{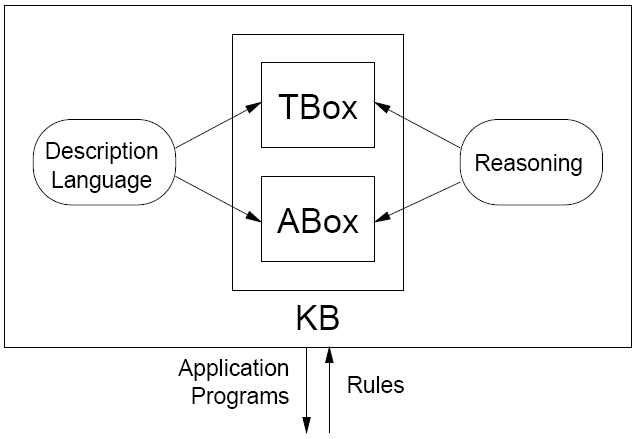}
\caption{The architecture of Knowledge Representation System based on Description Logics \cite{BCM03}}
\label{fig:DLSystemArchitecture}
\end{figure}

Descriptions of relationships (concepts and roles) and their combinations can be used to build Description Logics Knowledge Representations Systems (DLKRSs). Reasoning services of such systems are divided into 2 parts: \textit{terminological} component (called TBox) and an \textit{assertional} component (called ABox). Such a hybrid architecture was pioneered by the KRYPTON system, and has been adopted by many DLKRSs. It is shown on the Figure \ref{fig:DLSystemArchitecture}.

\newpage

\noindent TBox introduces the terminology which means the vocabulary of application domain. The vocabulary consist of \textit{concepts} (set of individuals) and \textit{roles} (binary relationships between individuals). TBox can also be used for creating names (abbreviations) for complex descriptions. ABox contains assertions about named individuals in terms of this vocabulary. When it comes to searching analogies for understanding what TBox and ABox component is, we can collate TBox part with database \textit{schema}, and ABox with database \textit{data}.

DL systems, besides storing terminologies and assertions, also provide reasoning mechanisms. Hence, TBox (term classifier) reasons about concepts and roles descriptions and their relationships, while ABox reasons about individuals and their relationships, e.g.
\begin{itemize}
    \item TBox:

\[
\begin{array}{lcl} 
	\mathit{GoodCPU}    & \equiv    & \mathit{Thing} \sqcap~ (\exists \mathit{isMadeOf}.\mathit{Metalloid}) \\ 
                        &           & \sqcap~ (\exists \mathit{isCreatedBy}.\mathit{ChipManufacturer}) \\
                        &           & \sqcap~ (\forall \mathit{hasFeature}.(\geq 4~ \mathit{hasCore} \sqcup \lnot \mathit{x86Arch})) \\
                        &           & \sqcap~ \exists \mathit{hasFeature}.\top
\end{array}
\]

In one of its simplest form TBox introduces abbreviation for a complex description. The abbreviation here is $\mathit{GoodCPU}$, which indicates something described by following statement: "a thing made of semi metal, which is created by chip maker and all of its features are either more than 3 cores or not x86 architecture".

\newpage
    \item ABox:

\[
\begin{array}{ll} 
    \mathit{GoodCPU}(\mathit{Itanium})              & \mathit{Metalloid}(\mathit{Silicon}) \\
    \mathit{ChipManufacturer}(\mathit{Intel Corp.}) & \mathit{hasFeature}(\mathit{Itanium}, \mathit{IA64}) \\
    \mathit{hasCore}(\mathit{Itanium}, 12)          & \lnot \mathit{x86Arch}(\mathit{IA64})
\end{array}
\]

We can see descriptions of specific situation, where properties are asserted for some individuals. $\mathit{GoodCPU}$ indicates the "Itanium chip made of Silicon by Intel Corporation, which has 12 cores based on IA64 architecture (not backward compatible with old x86)".

\end{itemize}

\noindent Reasoning about terminology (TBox) provides \textbf{satisfiability} (i.e. non-contradictory) and \textbf{subsumption} services. Satisfiability is used for finding an interpretation that makes the formula true and subsumption is used for organizing the concept hierarchy according to their generality (ordering of concepts based on subsumption relation). 

Reasoning about assertions (ABox) provides \textbf{consistency} and \textbf{instantiation} mechanisms. A theory is consistent when it does not contain contradiction. In the semantic terms it means, that it has a model. Instantiation is used to perform \textit{realization} and \textit{retrieval}. Realization means computing the most specific concepts which an individual instantiates while retrieval means computing the individuals which are instances of a given concept. 

Satisfiability checks of descriptions while consistency checks of sets of assertions to determine whether a knowledge base is meaningful or not.

On the start of typical DLKRS application, TBox reasoning services are invoked to ensure that all defined concepts are satisfiable, i.e. are not subsumed by the bottom concept, which is always interpreted as the empty set. Moreover, subsumption hierarchy is computed. This hierarchy would then be inspected to make sure that it coincides with the intention of the modeler. Next it comes to ABox, which first check for its consistency with the TBox and then, for example, compute the most specific concepts that each individual is an instance of (known as realizing the ABox).

\subsection{Description languages}
\label{sub:DLs}

As it was mentioned at the beginning, elementary descriptions are the expressions built from \textit{atomic concepts} (unary predicates) and \textit{atomic roles} (binary predicates). Complex descriptions are created from elementary descriptions inductively by using of \textit{concept constructors}.

Description languages are distinguished by the constructors they provide. The language $\mathcal{AL}$ (Attributive Language) is a minimal language of practical interest. The other languages of this family are extensions of $\mathcal{AL}$, i.e. $\mathcal{ALC}$ (Section \ref{sss:ALC}) is obtained from $\mathcal{AL}$ by adding the complement operator $(\lnot)$, $\mathcal{ALE}$ is obtained from $\mathcal{AL}$ by adding existential restrictions $(\exists R.C)$, etc. \cite{BCM03, HLP08}. Before moving into details, we will introduce conventional notation in DL:

\newpage

\begin{table}[htp]
\centering
\begin{tabular}{ |c|l| }
    \hline
    \textbf{Symbol} & \multicolumn{1}{c|}{\textbf{Description}} \\ \hline
    $\top$          & all concept names \\ \hline
    $\bot$          & empty  concept \\ \hline
    $\sqcap$        & intersection or conjunction of concepts \\ \hline
    $\sqcup$        & union or disjunction of concepts \\ \hline
    $\lnot$         & negation  or complement of concepts \\ \hline
    $\forall$       & universal restriction \\ \hline
    $\exists$       & existential restriction \\ \hline
    $\sqsubseteq$   & concept inclusion \\ \hline
    $\equiv$        & concept equivalence \\ \hline
    $\doteq$        & concept definition \\ \hline
    $:$             & concept/role assertion \\ \hline
\end{tabular}
\caption{DL notation}
\end{table} 

\noindent Let A and B be atomic concepts, R be atomic role, and C and D be concept descriptions. Based on that abstract notation, concept descriptions in $\mathcal{AL}$ language are formed by following syntax rule \cite{BCM03}:

\smallskip

\[
\begin{array}{lcll} 
    C,D & \to   & A~|               & (atomic~ concept) \\ 
        & 		& \top~|            & (universal~ concept) \\
        & 		& \bot~|            & (bottom~ concept) \\
        &		& \lnot A~|         & (atomic~ negation) \\
        &		& C \sqcap D~|      & (intersection) \\
        &		& \forall R.C~|     & (value~ restriction) \\
        &		& \exists R.\top~|  & (limited~ existential~ quantification)
\end{array}
\]

\smallskip

In this section a basic example of what can be expressed using $\mathcal{AL}$ is shown. Let's assume $\mathbf{Computer}$ and $\mathbf{Notebook}$ are atomic concepts. Based on that $\mathbf{Computer} \sqcup \mathbf{Notebook}$ is an $\mathcal{AL}-concept$ describing computers which are notebooks. In analogical way $\mathbf{Computer} \sqcup \lnot\mathbf{Notebook}$ describes concept of computers which are not notebooks (they can be calculators, supercomputers, etc.). Suppose $\mathbf{hasConnection}$ is an atomic role - we can describe the concept $\mathbf{Computer} \sqcap \exists\mathbf{hasConnection}.\top$ as indicating computers which have network connections. $\mathbf{Computer} \sqcap \forall\mathbf{hasConnection}.\mathbf{Notebook}$ describes such computers which are connected only to other notebooks. Concept of computers without any connections is $\mathbf{Computer} \sqcap \forall\mathbf{hasConnection}.\bot$.

When it comes to define semantics of $\mathcal{AL}-concepts$ we consider interpretation $\mathcal{I}$. The interpretation consists of a non-empty set $\Delta^{\mathcal{I}}$ (the domain of the interpretation) and an interpretation function, which assigns to every atomic concept $A$ a set $A^{\mathcal{I}} \subseteq \Delta^{\mathcal{I}}$ and to every atomic role $R$ a binary relation $R^{\mathcal{I}} \subseteq \Delta^{\mathcal{I}} \times \Delta^{\mathcal{I}}$. The interpretation function is extended to concept descriptions by the following inductive definitions \cite{BCM03}:

\newpage

\[
\begin{array}{rcl}
    \top^{\mathcal{I}}              & = & \Delta^{\mathcal{I}} \\
    \bot^{\mathcal{I}}              & = & \emptyset \\
    (\lnot A)^{\mathcal{I}}         & = & \Delta^{\mathcal{I}}\setminus A^{\mathcal{I}} \\
    (C\sqcap D)^{\mathcal{I}}       & = & C^{\mathcal{I}}\cap D^{\mathcal{I}} \\
    (\forall R.C)^{\mathcal{I}}     & = & \{a \in \Delta^{\mathcal{I}}~|~\forall b.~(a,b) \in R^{\mathcal{I}} \to b \in C^{\mathcal{I}}\} \\
    (\exists R.\top)^{\mathcal{I}}  & = & \{a \in \Delta^{\mathcal{I}}~|~\exists b.~(a,b) \in R^{\mathcal{I}}\}
\end{array}
\]

\smallskip

Two concepts $C$, $D$ are equivalent ($C \equiv D$), if $C^{\mathcal{I}} = D^{\mathcal{I}}$ for all interpretations $\mathcal{I}$, e.g. $\forall hasConnection.Notebook \sqcap \forall hasConnection.Ebook$ is equivalent to $\forall hasConnection.(Notebook \sqcup Ebook)$.

\subsubsection{Syntax and semantics of \texorpdfstring{$\mathcal{ALC}$}{ALC}} 
\label{sss:ALC}

The basic DL $\mathcal{ALC}$ stands for \textbf{Attributive Concept Language with Complements}. It was introduced by Manfred Schmidt-Schauß and Gert Smolka in 1991. $\mathcal{ALC}$ is obtained from $\mathcal{AL}$ by adding the complement operator $(\lnot)$. Below, there are definitions of syntax and semantics of $\mathcal{ALC}$ \cite{HLP08}:

\bigskip

\begin{definition}\label{alcSyntax} ($\mathcal{ALC}$ syntax). Let $N_{C}$ be a set of concept names and $N_{R}$ be a set of role names. The set of $\mathcal{ALC}$-concept descriptions is the smallest set such that

\begin{enumerate}
    \item $\top$, $\bot$, and every concept name $A\in N_{C}$ is an $\mathcal{ALC}$-concept description,
    \item if $C$ and $D$ are $\mathcal{ALC}$-concept descriptions and $r\in N^{R}$, then $C \sqcap D$, $C \sqcup D$, $\lnot C$, $\forall r.C$, and $\exists r.C$ are $\mathcal{ALC}$-concept descriptions.
\end{enumerate}

\noindent In the following, we will often use "$\mathcal{ALC}$-concept" instead of "$\mathcal{ALC}$-concept description". The semantics of $\mathcal{ALC}$ (and of DLs in general) is given in terms of interpretations.
\end{definition}

\medskip

\begin{definition}\label{alcSemantics} ($\mathcal{ALC}$ semantics). An interpretation $I = (\Delta^{\mathcal{I}},\cdot^{I})$ consists of a nonempty set $\Delta^{\mathcal{I}}$, called the domain of $\mathcal{I}$, and a function $\cdot^{\mathcal{I}}$ that maps every $\mathcal{ALC}$-concept to a subset of $\Delta^{\mathcal{I}}$, and every role name to a subset of $\Delta^{\mathcal{I}} \times \Delta^{\mathcal{I}}$ such that, for all $\mathcal{ALC}$-concepts $C$,$D$ and all role names $r$,

\smallskip

\[
\begin{array}{rcl}
    \top^{\mathcal{I}}          & =	& \Delta^{\mathcal{I}} \\
    \bot^{\mathcal{I}}          & =	& \emptyset \\
    (C\sqcap D)^{\mathcal{I}}   & =	& C^{\mathcal{I}}\cap D^{\mathcal{I}} \\
    (C\sqcup D)^{\mathcal{I}}   & =	& C^{\mathcal{I}}\cup D^{\mathcal{I}} \\
    (\lnot C)^{\mathcal{I}}     & =	& \Delta^{\mathcal{I}}\setminus C^{\mathcal{I}} \\
    (\forall r.C)^{\mathcal{I}} & =	& \{x \in \Delta^{\mathcal{I}}~|~ For~ all~ y \in \Delta^{\mathcal{I}}~ if~<x,y> \in r^{\mathcal{I}},~ then~ y \in C^{\mathcal{I}}\} \\
    (\exists r.C)^{\mathcal{I}} & =	& \{x \in \Delta^{\mathcal{I}}~|~There~ is~ some~ y \in \Delta^{\mathcal{I}}~ with~ <x,y> \in r^{\mathcal{I}}~ and~ y \in C^{\mathcal{I}}\}
\end{array}
\]

\smallskip

\noindent We say that $C^{\mathcal{I}}~ (r^{\mathcal{I}})$ is the extension of the concept $C$ (role name $r$) in the interpretation $\mathcal{I}$. If $x\in C^{\mathcal{I}}$, then we say that $x$ is an instance of $C$ in $\mathcal{I}$.
\end{definition}

\section{OWL}
\label{sec:owl}

\subsection{Introduction}
\label{sub:owlIntroduction}

The OWL Web Ontology Language (OWL), is an ontology language for the Semantic Web with formally defined meaning. It was released in February 2004 as a W3C recommendation. OWL lays on top of RDF and RDFS and comes with a larger vocabulary and stronger syntax. All of them have similar foundations, but OWL is a stronger language with greater machine interpretability than RDF. It can be used to define classes and properties like RDFS, but contains additional set of constructs, which gives to it more expressive power. OWL 2 ontologies provide classes, properties, individuals, and data values and are stored as Semantic Web documents. OWL ontologies can be used along with information written in RDF, and OWL ontologies themselves are primarily exchanged as RDF documents. \textit{"OWL's expressivity is sufficient to cover most of the well-known Description Logic formalisms"} \cite{HLP08, W3COWL}.

\subsection{DLs and OWL}
\label{sub:dlsAndOwl}

OWL is DL-based ontology language. Because the semantics of OWL (Lite and DL) \textit{"can be defined via a translation into an expressive DL"} \cite{HLP08}, implemented DL reasoners can be used for reasoning tasks in applications built on OWL. Analogies in the naming conventions between DL and OWL, are shown in Table \ref{tab:naming}.

\medskip

\begin{table}[htp]
\centering
\begin{tabular}{ |>{\centering\arraybackslash}m{2cm}|>{\centering\arraybackslash}m{2cm}| }
    \hline
    \multicolumn{1}{|c|}{\textbf{OWL}}  &	\multicolumn{1}{c|}{\textbf{DL}} \\ \hline
    class                               &	concept \\ \hline
    object                              &	individual \\ \hline
    property                            &	role \\ \hline
\end{tabular}
\caption{OWL DL naming synonyms}
\label{tab:naming}
\end{table}

\newpage

\noindent An OWL ontology describes the domain in the terms of classes (concepts), objects (individuals) and properties (roles). OWL classes can be created from basic classes or properties, by variety of constructors. The constructors supported by OWL, are shown in Table \ref{tab:owlConstructors}. An ontology consists of groups of axioms. Axioms are used for making the assertions, e.g. assertions of subsumption relationships between classes or properties. They are presented in Table \ref{tab:owlAxioms}.

\bigskip

\begin{table}[htp]
\centering
\begin{tabular}{ |>{\tt}l|l|l| }
    \hline
    \multicolumn{1}{|c|}{\textbf{Constructor}}  & \multicolumn{1}{c|}{\textbf{DL Syntax}}   & \multicolumn{1}{c|}{\textbf{Example}} \\ \hline
    intersectionOf                              & $C_{1}\sqcap\cdots\sqcap C_{n}$           & $Human \sqcap Male$ \\ \hline
    unionOf                                     & $C_{1}\sqcup\cdots\sqcup C_{n}$           & $\mathit{Doctor} \sqcup \mathit{Lawyer}$ \\ \hline
    complementOf                                & $\lnot C$                                 & $\lnot \mathit{Male}$ \\ \hline
    oneOf                                       & ${x_{1}\cdots x_{2}}$                     & ${\mathit{john},\mathit{mary}}$ \\ \hline
    allValuesFrom                               & $\forall P.C$                             & $\forall \mathit{hasChild}.\mathit{Doctor}$ \\ \hline
    someValuesFrom                              & $\exists r.C$                             & $\exists \mathit{hasChild}.\mathit{Lawyer}$ \\ \hline
    hasValue                                    & $\exists r.\{x\}$                         & $\exists \mathit{citizensOf}.\{\mathit{USA}\}$ \\ \hline
    minCardinality                              & $(\geq nr)$                               & $(\geq 2~\mathit{hasChild})$ \\ \hline
    maxCardinality                              & $(\leq nr)$                               & $(\leq 2~\mathit{hasChild})$ \\ \hline
    inverseOf                                   & $r^{-}$                                   & $\mathit{hasChild}^{-}$ \\ \hline
\end{tabular}
\caption{OWL constructors \cite{HLP08}}
\label{tab:owlConstructors}
\end{table}

\begin{table}[htp]
\centering
\begin{tabular}{ |>{\tt}l|l|l| }
    \hline
    \multicolumn{1}{|c|}{\textbf{Axiom}}    & \multicolumn{1}{c|}{\textbf{DL Syntax}}   & \multicolumn{1}{c|}{\textbf{Example}} \\ \hline
    subClassOf                              & $C_{1} \sqsubseteq C_{2}$                 & $\mathit{Human} \sqsubseteq \mathit{Animal} \sqcap \mathit{Biped}$ \\ \hline
    equivalentClass                         & $C_{1} \equiv C_{2}$                      & $\mathit{Man} \equiv \mathit{Human} \sqcap \mathit{Male}$ \\ \hline
    subPropertyOf                           & $P_{1} \sqsubseteq P_{2}$                 & $\mathit{hasDaughter} \sqsubseteq \mathit{hasChild}$ \\ \hline
    equivalentProperty                      & $P_{1} \equiv P_{2}$                      & $\mathit{cost} \equiv \mathit{price}$ \\ \hline
    disjointWith                            & $C_{1} \sqsubseteq \lnot C_{2}$           & $\mathit{Male} \sqsubseteq \lnot \mathit{Female}$ \\ \hline
    sameAs                                  & ${x_{1}} \equiv {x_{2}}$                  & $\mathit{President\_Bush} \equiv \mathit{G\_W\_Bush}$ \\ \hline
    differentFrom                           & ${x_{1}} \sqsubseteq \lnot {x_{2}}$       & $\mathit{john} \sqsubseteq \lnot \mathit{peter}$ \\ \hline
    TransitiveProperty                      & $P~transitive~role$                       & $\mathit{hasAncestor}~is~a~transitive~role$ \\ \hline
    FunctionalProperty                      & $T \sqsubseteq (\leq 1~P)$                & $T \sqsubseteq (\leq 1~\mathit{hasMother})$ \\ \hline
    InverseFunctionalProperty               & $T \sqsubseteq (\leq 1~P^{-})$            & $T \sqsubseteq (\leq 1~\mathit{isMotherOf}^{-})$ \\ \hline
    SymmetricProperty                       & $P \equiv P^{-}$                          & $\mathit{isSiblingOf} \equiv \mathit{isSiblingOf}^{-}$ \\ \hline
\end{tabular}
\caption{OWL axioms \cite{HLP08}}
\label{tab:owlAxioms}
\end{table}

\newpage

\noindent Based on the DL syntax shown in tables above, we can construct OWL equivalents by serializing them to XML, e.g.

\[
\mathit{Human} \sqcap \mathit{Male}
\]

{\tt \small
\begin{verbatim}
<owl:intersectionOf rdf:parseType="Collection">
   <rdf:Description rdf:about="#Human"/>
   <rdf:Description rdf:about="#Male"/>
</owl:intersectionOf>
\end{verbatim}
}

\[
\mathit{Doctor} \sqcup \mathit{Lawyer}
\]

{\tt \small
\begin{verbatim}
<owl:unionOf rdf:parseType="Collection">
   <rdf:Description rdf:about="#Doctor"/>
   <rdf:Description rdf:about="#Lawyer"/>
</owl:unionOf>
\end{verbatim}
}

\[
\exists \mathit{hasChild}.\mathit{Lawyer}
\]

{\tt \small
\begin{verbatim}
<owl:Restriction>
   <owl:onProperty rdf:resource="#hasChild"/>
   <owl:someValuesFrom rdf:resource="#Lawyer"/>
</owl:Restriction>
\end{verbatim}
}

\[
\exists \mathit{citizensOf}.\{\mathit{USA}\}
\]

{\tt \small
\begin{verbatim}
<owl:Restriction>
   <owl:onProperty rdf:resource="#isCitizenOf"/>
   <owl:hasValue rdf:resource="#USA"/>
</owl:Restriction>
\end{verbatim}
}

\bigskip

\noindent and last, slightly more complicated example:

\[
\begin{array}{lcl} 
    \mathit{GoodCPU}    & \equiv    & \mathit{Thing} \sqcap~ (\exists \mathit{isMadeOf}.\mathit{Metalloid}) \\ 
                        &           & \sqcap~ (\exists \mathit{isCreatedBy}.\mathit{ChipManufacturer}) \\
                        &           & \sqcap~ (\forall \mathit{hasFeature}.(\geq 4~ \mathit{hasCore} \sqcup \lnot \mathit{x86Arch}))
\end{array}
\]

\newpage

{\tt \footnotesize
\begin{verbatim}
<owl:Class rdf:about="#GoodCPU">
   <owl:equivalentClass>
      <owl:Class>
         <owl:intersectionOf rdf:parseType="Collection">
            <rdf:Description rdf:about="&owl;Thing"/>
            <owl:Restriction>
               <owl:onProperty rdf:resource="#createdBy"/>
               <owl:someValuesFrom rdf:resource="#ChipManufacturer"/>
            </owl:Restriction>
            <owl:Restriction>
               <owl:onProperty rdf:resource="#madeOf"/>
               <owl:someValuesFrom rdf:resource="#Metalloid"/>
            </owl:Restriction>
            <owl:Restriction>
               <owl:onProperty rdf:resource="#hasFeature"/>
               <owl:allValuesFrom>
                  <owl:Class>
                     <owl:unionOf rdf:parseType="Collection">
                        <owl:Class>
                           <owl:complementOf rdf:resource="#x86Arch"/>
                        </owl:Class>
                        <owl:Restriction>
                           <owl:onProperty rdf:resource="#hasCore"/>
                           <owl:someValuesFrom>
                              <rdf:Description>
                                 <rdf:type rdf:resource="&rdfs;Datatype"/>
                                 <owl:onDatatype rdf:resource="&xsd;integer"/>
                                 <owl:withRestrictions rdf:parseType="Collection">
                                    <rdf:Description>
                                       <xsd:minInclusive 
                                            rdf:datatype="&xsd;integer">4
                                       </xsd:minInclusive>
                                    </rdf:Description>
                                 </owl:withRestrictions>
                              </rdf:Description>
                           </owl:someValuesFrom>
                        </owl:Restriction>
                     </owl:unionOf>
                  </owl:Class>
               </owl:allValuesFrom>
            </owl:Restriction>
         </owl:intersectionOf>
      </owl:Class>
   </owl:equivalentClass>
</owl:Class>
\end{verbatim}
}

\subsubsection{OWL and reasoners}
\label{sub:owlAndReasoners}

Possibility of using DL reasoning services in OWL-based applications was one of the reasons why OWL is based on DL. \textit{"OWL DL has a formal model-theoretic semantics providing a rigorous and provably decidable semantics for the language"} \cite{HLP08}. The decidability ensures, that consistency of OWL DL ontology can be checked by complete DL reasoners. Reasoners can be also used to infer information from asserted facts. 

Popular reasoners in the OWL community are: FaCT, FaCT++, Pellet, RACER, KAON2 or HermiT (see Section \ref{sss:hermiT}). These systems provide reasoning support for a set of tools designed for ontology creation and maintenance which has already been created. There are: Protégé (used for creation of traffic danger ontology described in Chapter \ref{cha:trafficDangerOntology}), Swoop, OilEd and TopBraid Composer.

Increasing number of tools designed for OWL is both a cause and a motivation for the community, to develop ontologies in many various fields, not only in the scope of the Semantic Web. There are areas like: biology, medicine, geography,
geology, astronomy, agriculture or defense, in which ontologies become adopted \cite{HLP08}.

The specialists from IBM departments in China and USA and Department of Biomedical Informatics in Columbia University, have prepared a report about their experiences with Semantic Web applications \cite{UCEReport}. They were working through large terminology called MED (Medical Entities Dictionary) used at the Columbia Presbyterian Medical Center. MED which previously had been using frame-based logic, has been transformed into OWL ontology. After transformation DL subsumption reasoning service was used to classify that ontology, which has revealed many modeling errors. But, as it is said in the report, \textit{"the important result here is not that we identified these modeling errors due to the increased expressivity of DL. More important is our finding that the missed subsumptions could have cost the hospital many missing results in various decision support and infection control systems that routinely use MED to screen patients."}

\subsection{OWL sublanguages}
\label{sub:owlSublanguages}

W3C specification defines 3 variants of OWL. These sublanguages: OWL Lite, OWL DL, and OWL Full provides different level of expressiveness (increasingly in mentioned order). Each of them contains extensions to its predecessor. Based on scope and complexity of the application domain, appropriate version should be chosen for description of the domain. More complex version gives more freedom in domain modeling. The complexity however affects the cost of a learning, which results in higher learning curve. We have 3 sublanguages inside OWL:
\begin{itemize}
    \setlength{\itemsep}{0cm}
    \setlength{\parskip}{0cm}

    \item \textbf{OWL Lite}, the simplest version, useful enough to support creation of classification hierarchies with simple constraints. It is not widely used and acts as the entry point for Semantic Web application developers.
    \item \textbf{OWL DL}, the most widely used version, includes OWL Lite. It was designed to provide maximum expressiveness possible while retaining computational completeness, decidability, and automated reasoning. That is why OWL DL takes the advantage of reasoning services provided for description logics (see Section \ref{sub:owlAndReasoners}).
    \item \textbf{OWL Full}, the most powerful in expressiveness but also the heaviest in the meaning of a learning cost. Includes OWL DL. Provides different semantic than predecessors.
\end{itemize}

\subsection{OWL 2 versus OWL 1}
\label{sub:owl2vsowl1}

\begin{figure}[htp]
\centering
\includegraphics[scale=0.5]{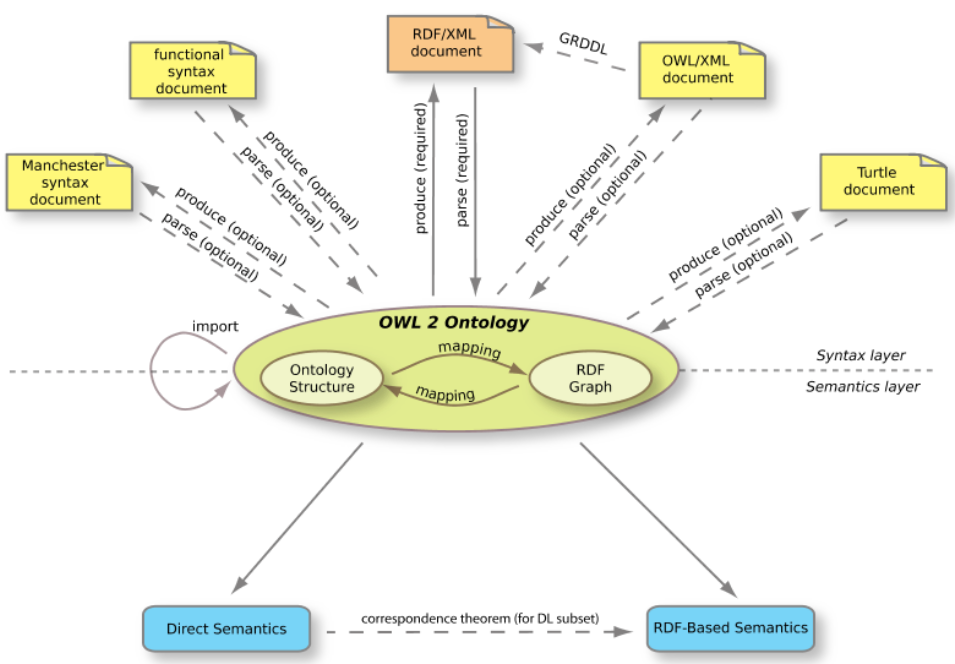}
\caption{The structure of OWL 2 \cite{W3COWL}}
\label{fig:owl2Structure}
\end{figure}

OWL 2 was announced by W3C working group on 27 October 2009, as the extension for previous versions (OWL and OWL 1.1). Like OWL 1, OWL 2 is designed to allow development of ontologies and to simplify this process. The second common design goal is to facilitate sharing ontologies via the web, with the aim of making web content more accessible to machines. Besides, OWL 2 has a very similar overall structure to OWL 1. Figure \ref{fig:owl2Structure} shows main building blocks of OWL 2 and relations between them. Despite changes in naming, almost all the building blocks of OWL 2 were present in OWL 1. All OWL 1 ontologies are valid OWL 2 ontologies.

\medskip

\newpage

\noindent Looking at the structure diagram, we can see that abstract notion of the ontology (abstract structure or RDF graph) is placed in the middle of it. At the top there are various serialization syntaxes for ontologies. At the bottom there are the two semantic specifications. These specifications define the meaning of OWL 2 ontologies. For most of users from the community, only one syntax and one semantic is sufficient. OWL 2 adds some new functionality but still remains compatible with OWL 1. Some of the new features include:
\begin{itemize}
    \setlength{\itemsep}{0cm}
    \setlength{\parskip}{0cm}

    \item keys,
    \item property chains,
    \item richer datatypes, data ranges,
    \item qualified cardinality restrictions,
    \item asymmetric, reflexive, and disjoint properties,
    \item enhanced annotation capabilities.
\end{itemize}

\noindent OWL 2 also defines three new profiles (sublanguages) (OWL 2 EL, OWL 2 QL, OWL 2 RL) and a new syntax (Manchester Syntax). In addition, the set of RDF Graphs that can be handled by DL reasoners is slightly larger in OWL 2 than in its predecessor. It is connected with the fact that some of the restrictions applicable to OWL DL have been relaxed in the newer OWL version \cite{W3COWL}.

\chapter{Traffic danger ontology}
\label{cha:trafficDangerOntology}

\section{Introduction}
\label{sec:introductionTochapter3}

An ontology is described as formal representation of the knowledge about some domain, by a set of concepts within the domain, and the relationships between those concepts. It can be used either to define the domain, or to reason the properties of that domain. One of formal definitions of ontology is \textit{"formal, explicit specification of a shared conceptualization"} \cite{Gru93}. An ontology provides a shared vocabulary, which can be used to model a domain - that is, the type of objects and/or concepts that exist, and their properties and relations \cite{Arv08}.

Ontologies are used in artificial intelligence, Semantic Web, systems engineering, software engineering, biomedical informatics, etc. There are used as a form of knowledge representation about some part of world.

\begin{figure}[htp]
\centering
\includegraphics[scale=0.7]{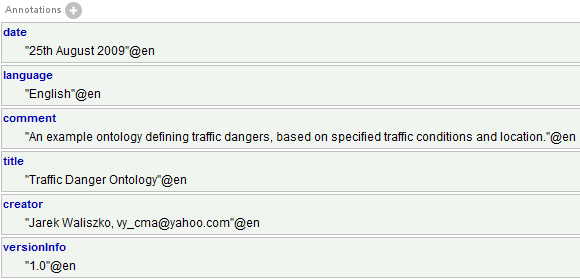}
\caption{Traffic danger ontology annotations in Protégé}
\label{fig:annotations}
\end{figure}

Ontologies may be represented by a lot of standards. Different ontology languages provide different facilities. The most recent development in standard ontology languages is OWL from the World Wide Web Consortium (W3C). One of the last tools for developing ontologies is Protégé \cite{OWLGuide}. Traffic danger ontology is developed using that tool. In Figure \ref{fig:annotations} there are annotations of that ontology, made in Protégé.

\medskip

\newpage

\begin{framed}
\noindent For the sake of clearness - almost all figures in this chapter refers to the traffic danger ontology and are all screenshots from Protégé tool. That assumption let avoid repeating under all figures the same supplement description (that figures are connected to traffic danger ontology, and are made in Protégé).
\end{framed}

\section{Knowledge engineering methodology for development of traffic danger concept}
\label{sec:knowledgeEngineering}

The real world provides a set of methodologies for developing an ontology, but there is no the fixed, best one. The domain experts can choose how to build ontologies in the way is best for them. While developing traffic danger ontology, the following fundamental rules described in \cite{OntDev101} were taken into consideration:
\begin{itemize}
    \item There is no one correct way to model a domain - there are always viable alternatives. The best solution almost always depends on the application that you have in mind and the extensions that you anticipate.
    \item Ontology development is necessarily an iterative process.
    \item Concepts in the ontology should be close to objects (physical or logical) and relationships in your domain of interest. These are most likely to be nouns (objects) or verbs (relationships) in sentences that describe your domain.
\end{itemize}

\noindent While developing an ontology it is important to have in mind that ontology is a model of real domain, so it must reflect reality. Details connected with that part of reality are going to be clearer and more crystallized down the road of building the ontology. After the initial version of the ontology, it is necessary to revise and rethink the design and repeat such iterative process to get the model that fulfill preferences and will be ready for giving answers for specific questions. Developing an ontology is not an aim on its own, usually it is a part of more complex software architecture which is driven by such formal domain model. Ontology should be applicable for the system it is going to cooperate. 

Methodology outlined above derives from agile development practices based on short feedback loops, systematic tests and constant cooperation with domain experts.

Ontology designing process requires to determine its domain and scope. It should be the starting point of work. That part is definitely more crystallized after answering several helpful questions \cite{OntDev101}:
\begin{itemize}
    \setlength{\itemsep}{0cm}
    \setlength{\parskip}{0cm}

    \item What is the domain that the ontology will cover?
    \item For what we are going to use the ontology?
    \item For what types of questions the information in the ontology should provide answers?
    \item Who will use and maintain the ontology?
\end{itemize}

\noindent The domain of traffic danger ontology consists of: traffic conditions, locations where such conditions can occur and the actual threats (related to specified conditions). The ontology will be used as the pillar of web system, which can provide real time information for traffic users about dangers connected with various areas. The ontology will be used by the public, it means, all end users who interact with the system. Besides, ontology will be widely accessible for anyone, who can use it for developing their own ideas based on that concept, or extend/improve existing formalization of the domain. Maintain of ontology will be narrowed for those trusted users, who can provide coherent, reliable data about current conditions on the roads. When it comes to competency questions \cite{FoxGru} the ontology is going to answer, traffic danger ontology passed through the following sample ones:
\begin{itemize}
    \setlength{\itemsep}{0cm}
    \setlength{\parskip}{0cm}

    \item What are the traffic dangers on specific area?
    \item Are there any dangers connected with low friction on specific area?
    \item What are the subareas of specific location?
    \item What kind of dangers are connected with bad atmospheric conditions?
    \item Is there any danger connected with specific postal code on specific district?
    \item Are there any traffic conditions provided for specific location?
\end{itemize}

\noindent After that part, domain developer should have almost clear vision of what knowledge structure is going to be created. Searching for already defined ontologies, which could fit desired problem, or at least some of its parts, is a good starting point. 

When it comes to the traffic danger ontology development process, the primary research has resulted in the finding of geographical locations ontology. Unfortunately that information structure was too complex to be included into the prototype. After all, it was not an easy task to find any helpful topics on the web - at least not formalized in a way simple enough that could be useful for the needs of this thesis. Because of that, creation of new ontology has started from scratch.

The top-down development process has been chosen for ontology creation. Starting from the definition of the most general concepts in the domain, ontology construction has been carried on, down to the most detailed concepts. After one concept had been entirely built up, and has been sufficient for the current needs, development of the next one has been started in the similar top-down way.

After all parts of ontology had been built, analysis of the ontology was started. The analysis procedure could be compared to specific kind of debugging process. In that part, reasoner was used for verifying answers for competency questions. If some parts of ontology was not coherent with another one, or the ontology was not able to result in providing correct answers, the broken parts were rebuilt. It was an iterative process, which ended when all competency questions was covered. Some of the questions has changed during the time of development process, but the latest form of the ontology fulfills answers to all questions correctly.

\section{Ontology parts description}
\label{sec:description}

In this section we will take a look at short outline of components that builds the OWL ontologies. We can talk about Individuals, Properties and Classes.

\subsection{Individuals}
\label{sub:individuals}

\begin{figure}[htp]
\centering
\includegraphics[scale=0.7]{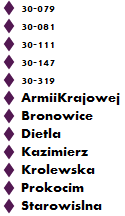}
\caption{Sample individuals}
\label{fig:individuals}
\end{figure}

The individuals could be treated as instances of classes. They represent the concrete objects, in our domain. Figure \ref{fig:individuals} shows sample individuals, used in ontology we are talking about. 

\begin{framed}
\noindent These sample individuals shown below, are not present in the pure ontology which describes the prototype concept of traffic danger. Individuals, and all links connected to them, are added to ontology after synchronization with database, which I'll explain in Chapter \ref{cha:trafficDangerWebSystem}.
\end{framed}

\smallskip

\newpage

\subsection{Properties}
\label{sub:properties}

\begin{figure}[htp]
\centering
\includegraphics[scale=0.7]{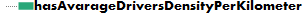}
\caption{Only one data property in our ontology}
\label{fig:dataProperties}
\end{figure}

\begin{figure}[htp]
\centering
\includegraphics[scale=0.7]{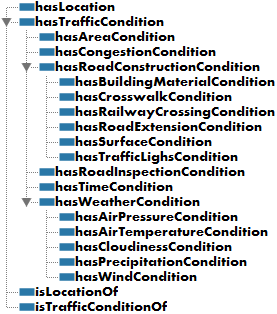}
\caption{Object properties}
\label{fig:objectProperties}
\end{figure}

Properties are binary relations between classes (or individuals). They account for linking individuals. In Figures \ref{fig:dataProperties} and \ref{fig:objectProperties} we can see respectively data and object properties used in traffic danger ontology.

\bigskip

\noindent Properties can have many features, e.g. they can have inverse properties, as discussed in the example below.

\bigskip

\begin{figure}[htp]
\centering
\includegraphics[scale=0.7]{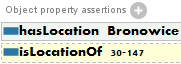}
\caption{Property assertions (for the street named \textbf{ArmiiKrajowej} individual)}
\label{fig:armiiKrajowej}
\end{figure}

\begin{figure}[htp]
\centering
\includegraphics[scale=0.7]{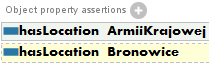}
\caption{Property assertions (for the postal code \textbf{30-147} individual)}
\label{fig:30-147}
\end{figure}

\noindent The property named \textbf{hasLocation} link the postal code individual \textbf{30-147} to the street individual called \textbf{ArmiiKrajowej} (Figure \ref{fig:armiiKrajowej}). The inverse property of \textbf{hasLocation} is \textbf{isLocationOf}. Because of the inversion, we can see, that the reasoner properly infer, that property \textbf{isLocationOf} link \textbf{ArmiiKrajowej} street to \textbf{30-147} postal code (Figure \ref{fig:30-147}). Inferred features are marked by light yellow background. The reasoning is based on description logics (DL), which will be explained later.

\medskip

\newpage

\subsection{Classes}
\label{sub:classes}

\begin{figure}[htp]
\centering
\includegraphics[scale=0.7]{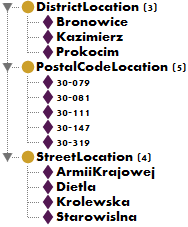}
\caption{Individuals by class}
\label{fig:individualsByClass}
\end{figure}

\begin{figure}[htp]
\centering
\begin{tabular}[t]{ll}
\multirow{2}{*}{\includegraphics[scale=0.7]{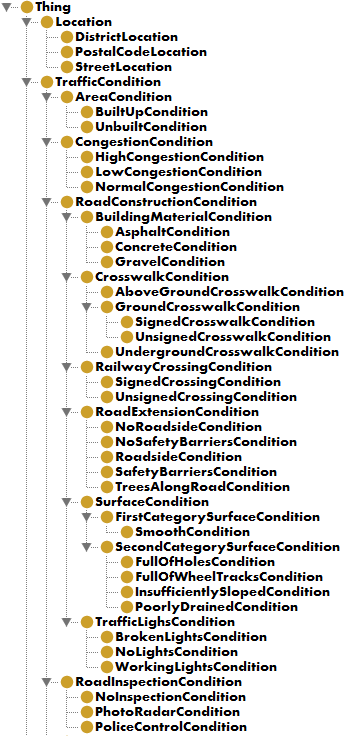}} & continuation... \vspace{0.6cm} \\
& \includegraphics[scale=0.7]{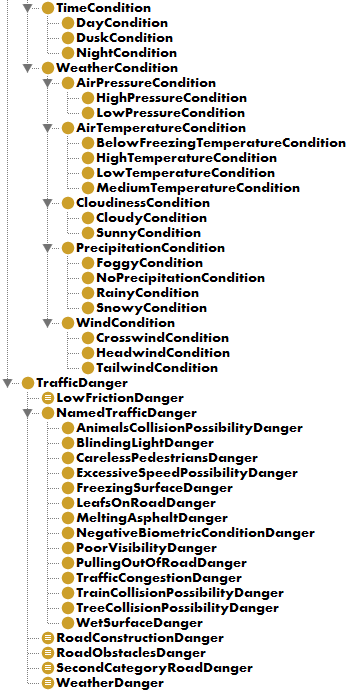}
\end{tabular}
\caption{Asserted class hierarchy}
\label{fig:assertedClassHierarchy}
\end{figure}

If we use analogy, the classes may be viewed as types in object oriented programming. They represent the concepts, in the other words, they define the abstracts. They also may be treated as sets of individuals, and contain all individuals in our domain of concept, that fill all requirements of membership. Classes are though described using formal, mathematical descriptions called description logics (OWL DL). These logics allow, to precisely certify, if certain subclasses or individuals are members of our superclass, or not. The process of deduction may be computed automatically using reasoners. That is one of the key features of OWL-DL. Figure \ref{fig:individualsByClass} shows sample individuals assigned to their classes. Figure \ref{fig:assertedClassHierarchy} shows all classes, traffic danger ontology is consisted of.

\medskip

\newpage

\section{Existential and universal restrictions}
\label{sec:existentialAndUniversalRestrictions}

This section considers the important issue of existential and universal restrictions in Protégé. As it was described in Sections \ref{sec:dl} and \ref{sec:owl}, existential restrictions are denoted in formal Description Logics (DL) Syntax by $\mathit{existential~ quantifier} (\exists)$, and in Ontology Web Language (OWL) by $\mathit{someValuesFrom}$. Universal restrictions on the other hand, are denoted in DL Syntax by $\mathit{universal~ quantifier} (\forall)$, and in OWL by $\mathit{allValuesFrom}$.
\begin{itemize}
    \item \textit{"Existential restrictions describe classes of individuals that participate in at least one relationship along a specified property to individuals that are members of a specified class"} \cite{OWLGuide}. 

In Protégé the keyword \textbf{some} is used to denote existential restrictions.
    \item \textit{"Universal restrictions describe classes of individuals that for a given property only have relationships along this property to individuals that are members of a specified class"} \cite{OWLGuide}.

In Protégé the keyword \textbf{only} is used  to denote universal restrictions.
\end{itemize}

\section{Open world assumption and closure axiom}
\label{sec:openWorldAssumptionAndClosureAxiom}

This section is directly connected with the reasoning domain (see Section \ref{sec:reasoning}). The reason of separation this field to its individual section is to emphasize its importance.

Open word assumption (OWA) is one of the foundations of reasoning mechanisms used in Description Logics and OWL. In other papers OWA can be referred as OWR which extends to open world reasoning.

The OWA denotes, that a fact which has not been stated to be true, cannot be assumed to be false. In the other words there is prohibited to assume that something does not exist until that fact is explicitly stated. Such an ambiguous situation simply means that the knowledge has not yet been added to the knowledge base.

In the case of described traffic ontology, that is stated that \textbf{PoorVisibilityDanger} has precipitation conditions that are kinds of \textbf{FoggyCondition}, \textbf{RainyCondition} and \textbf{SnowyCondition}. Because of the open world assumption, until we explicitly say that a \textbf{PoorVisibilityDanger} only has that kinds of conditions, it is assumed (by the reasoner) that a \textbf{PoorVisibilityDanger} could have other conditions such as \textbf{SunnyCondition}. To specify explicitly that a \textbf{BlindingLightDanger} has conditions that are kinds of \textbf{FoggyCondition}, \textbf{RainyCondition} and \textbf{SnowyCondition} only, a closure axiom has to be added on the \textbf{hasPrecipitationCondition} property.

A closure axiom on a property denotes that property can be filled by the specified set of fillers. The axiom consists of a universal restriction which has a filler that is the union of fillers that occur in the existential restrictions for the property.

\newpage

In traffic danger ontology for example, the closure axiom on the \textbf{hasPrecipitationCondition} for \textbf{PoorVisibilityDanger} defined by following existential restrictions:

\[
\begin{array}{l} 
\exists hasPrecipitationCondition. FoggyCondition \\ 
\exists hasPrecipitationCondition. RainyCondition \\
\exists hasPrecipitationCondition. SnowyCondition
\end{array}
\]

\smallskip

{\tt \small
\begin{verbatim}
<owl:Class rdf:about="#PoorVisibilityDanger">
   <rdfs:label xml:lang="en">PoorVisibilityDanger</rdfs:label>
   <rdfs:subClassOf>
      <owl:Restriction>
         <owl:onProperty rdf:resource="#hasPrecipitationCondition"/>
         <owl:someValuesFrom rdf:resource="#FoggyCondition"/>
      </owl:Restriction>
   </rdfs:subClassOf>
   <rdfs:subClassOf>
      <owl:Restriction>
         <owl:onProperty rdf:resource="#hasPrecipitationCondition"/>
         <owl:someValuesFrom rdf:resource="#RainyCondition"/>
      </owl:Restriction>
   </rdfs:subClassOf>
   <rdfs:subClassOf>
      <owl:Restriction>
         <owl:onProperty rdf:resource="#hasPrecipitationCondition"/>
         <owl:someValuesFrom rdf:resource="#SnowyCondition"/>
      </owl:Restriction>
   </rdfs:subClassOf>
</owl:Class>
\end{verbatim}
}

\medskip

\begin{figure}[htp]
\centering
\includegraphics[scale=0.7]{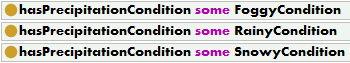}
\caption{Existential restrictions on \textbf{hasPrecipitationCondition} property}
\label{fig:existentialRestrictions}
\end{figure}

\newpage

\noindent is stated as a universal restriction, which acts along the \textbf{hasPrecipitationCondition} property, with a filler that is the union of \textbf{FoggyCondition}, \textbf{RainyCondition} and also \textbf{SnowyCondition}:

\[
\forall hasPrecipitationCondition. (FoggyCondition \sqcup RainyCondition \sqcup SnowyCondition)
\]

\smallskip

{\tt \small
\begin{verbatim}
<owl:Class rdf:about="#PoorVisibilityDanger">
   <rdfs:subClassOf>
      <owl:Restriction>
         <owl:onProperty rdf:resource="#hasPrecipitationCondition"/>
         <owl:allValuesFrom>
            <owl:Class>
               <owl:unionOf rdf:parseType="Collection">
                  <rdf:Description rdf:about="#FoggyCondition"/>
                  <rdf:Description rdf:about="#RainyCondition"/>
                  <rdf:Description rdf:about="#SnowyCondition"/>
               </owl:unionOf>
            </owl:Class>
         </owl:allValuesFrom>
      </owl:Restriction>
   </rdfs:subClassOf>
</owl:Class>
\end{verbatim}
}

\medskip

\begin{figure}[htp]
\centering
\includegraphics[scale=0.7]{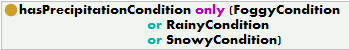}
\caption{Closure axiom on \textbf{hasPrecipitationCondition} property}
\label{fig:closureAxiom}
\end{figure}

\newpage

\section{Reasoning}
\label{sec:reasoning}

\subsection{Introduction}
\label{sub:introduction}

\begin{figure}[htp]
\centering
\includegraphics[scale=0.67]{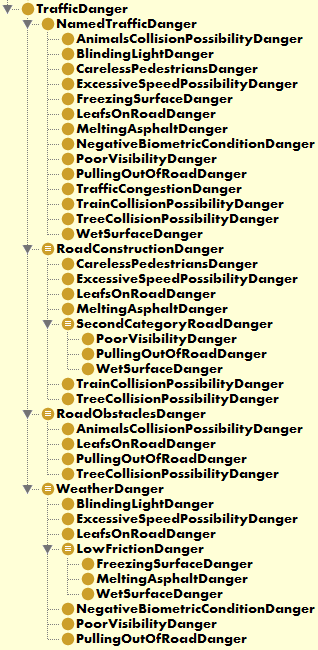}
\caption{Defined classes with inferred memberships from \textbf{NamedTrafficDanger} subclasses}
\label{fig:inferredClassHierarchy}
\end{figure}

Such classes that do not have any sets of necessary and sufficient conditions, but have only necessary conditions, are known as primitive classes. In Protégé they are marked by plain round yellow icon. Yet the classes marked by 3 horizontal white lines, are called defined classes. It means that such classes have at least one set of the necessary and sufficient conditions for the reasoner, to make assumptions based on their descriptions (to infer relationships). The reasoner is able to infer dependencies only for defined classes \cite{OWLGuide}. When we turn on the reasoner, the result of inferred classes will be like that presented in Figure \ref{fig:inferredClassHierarchy}.

\medskip

\newpage

\begin{figure}[htp]
\centering
\includegraphics[scale=0.7]{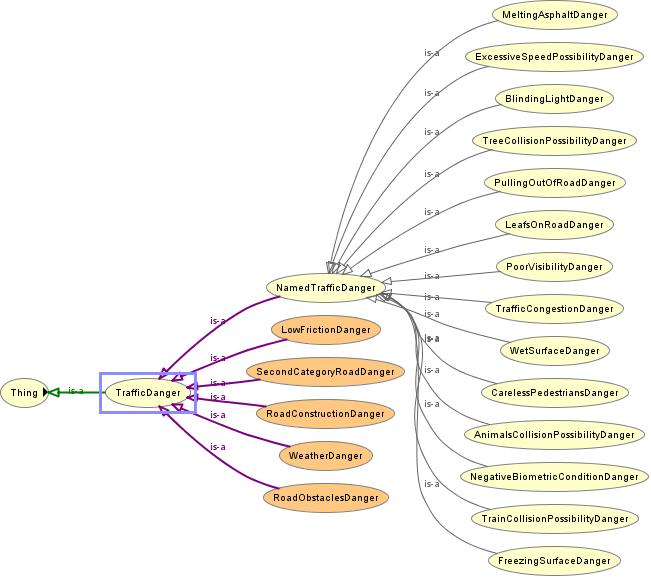}
\caption{Graphviz made view showing subclasses structure of \textbf{TrafficDanger} class}
\label{fig:trafficDangerOwlViz}
\end{figure}

\noindent Another view of this taxonomy can be generated using Graphviz tool. It is open source graph (network) visualization project from AT\&T Research. Graphviz is integrated with Protégé as a plugin called OWLViz. We can take a look of \textbf{TrafficDanger} class subclasses structure in Figure \ref{fig:trafficDangerOwlViz}.

\medskip

\begin{figure}[htp]
\centering
\includegraphics[scale=0.6]{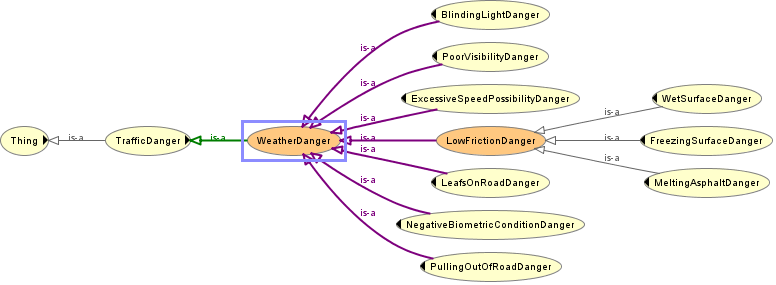}
\caption{Inferred members of \textbf{WeatherDanger} type}
\label{fig:weatherDangerInferredOwlViz}
\end{figure}

\noindent That orange ellipses shows the defined classes, which we were talking about earlier. The light yellow ellipses show primitive types. At the above picture we can see all subclasses of \textbf{NamedTrafficDanger} class. All of them have conditions based on which reasoner can make assumptions, and infer them to according defined classes. After inferring we can use OWLViz to show us the inferred relationships. Figure \ref{fig:weatherDangerInferredOwlViz} shows the inferred memberships of \textbf{WeatherDanger} class.

\newpage

\noindent The reasoner infer memberships of \textbf{WeatherDanger} type. We can see that \textbf{LowFrictionDanger} defined class is also subclass of \textbf{WeatherDanger}. Additionally that class has its own deducted members such as: \textbf{WetSurfaceDanger}, \textbf{FreezingSurfaceDanger} and \textbf{MeltingAsphaltDanger}.

How the classes were inferred, to be finally matched under appropriate supertypes? It is all because of description logics, the reasoning process of the reasoner engine is based on.

\subsection{First example}
\label{sub:firstExample}

\begin{figure}[htp]
\centering
\includegraphics[scale=0.7]{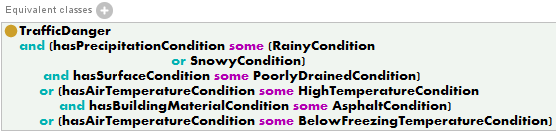}
\caption{OWL DL description of \textbf{LowFrictionDanger}}
\label{fig:lowFrictionDanger}
\end{figure}

In this example \textbf{LowFrictionDanger} deduction case is exercised. Below, in Figure \ref{fig:lowFrictionDanger}, we can see DL description, which tell us something about that class.

\medskip

\noindent Based on DL description, the subclasses tree is build. All we want to do is to classify specific dangers from \textbf{NamedTrafficDanger} type, to wider danger type called \textbf{LowFrictionDanger}. For that danger to occur, the following complex criteria must be fulfilled:

\newpage
\begin{enumerate}
    \setlength{\itemsep}{0cm}
    \setlength{\parskip}{0cm}

    \item there has to occur precipitation like rain or snow, and simultaneously the road has to have damaged system of draining water out from surface, or
    \item the air temperature has to be high while we are driving on asphalt road, or finally
    \item the temperature has to be below freezing, (because the surface is frozen and hence slippery)
\end{enumerate}

\begin{figure}[htp]
\centering
\includegraphics[scale=0.7]{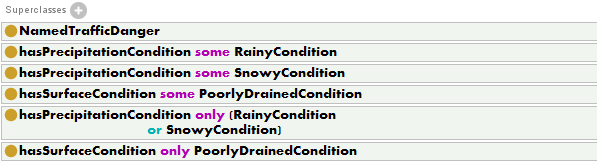}
\caption{\textbf{WetSurfaceDanger} concept superclasses set}
\label{fig:wetSurfaceDanger}
\end{figure}

\begin{figure}[htp]
\centering
\includegraphics[scale=0.7]{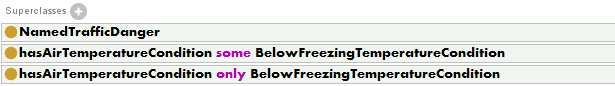}
\caption{\textbf{FreezingSurfaceDanger} concept superclasses set}
\label{fig:freezingSurfaceDanger}
\end{figure}

\begin{figure}[htp]
\centering
\includegraphics[scale=0.7]{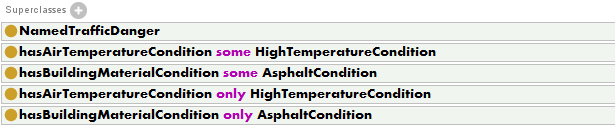}
\caption{\textbf{MeltingSurfaceAsphalt} concept superclasses set}
\label{fig:meltingAsphaltDanger}
\end{figure}

\noindent Because we know now (from Figure \ref{fig:weatherDangerInferredOwlViz}), the inferred members of \textbf{LowFrictionDanger} (\textbf{WetSurfaceDanger}, \textbf{FreezingSurfaceDanger} and \textbf{MeltingAsphaltDanger}), we should take a look of theirs DL descriptions. Figures \ref{fig:wetSurfaceDanger}, \ref{fig:freezingSurfaceDanger} and \ref{fig:meltingAsphaltDanger} show the dangers in mentioned order.

\medskip

\newpage

\begin{figure}[htp]
\centering
\includegraphics[scale=0.7]{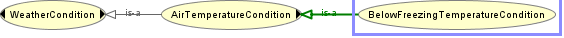}
\caption{Superclasses path of \textbf{BelowFreezingTemperatureCondition} class}
\label{fig:belowFreezingTemperatureConditionOWLViz}
\end{figure}

\noindent If we spent a while on analyzing the descriptions, the deduction which reasoner makes should by obvious and easy. Take a closer look at \textbf{FreezingSurfaceDanger} (Figure \ref{fig:freezingSurfaceDanger}). We can see that such a cold situation may occur only when the temperature is lower than 0. That is necessary condition that has to occur for the class, to be matched. \textbf{BelowFreezingTemperatureCondition} is defined as a subset of \textbf{AirTemperatureCondition}, and that one in turn is a subset of \textbf{WeatherCondition} (Figure \ref{fig:belowFreezingTemperatureConditionOWLViz}).

\medskip

\begin{figure}[htp]
\centering
\includegraphics[scale=0.7]{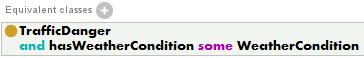}
\caption{\textbf{WeatherDanger} concept superclasses set}
\label{fig:weatherDanger}
\end{figure}

\noindent The description of \textbf{WeatherDanger} class (Figure \ref{fig:weatherDanger}) tells, that such kinds of dangers can be caused by inappropriate weather conditions.

\medskip

\noindent Because \textbf{BelowFreezingTemperatureCondition} is weather condition (subset of \textbf{WeatherCondition} type), \textbf{FreezingSurfaceDanger} is a member of \textbf{WeatherDanger} type.

\subsection{Second example}
\label{sub:secondExample}

\begin{figure}[htp]
\centering
\includegraphics[scale=0.7]{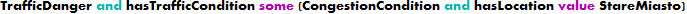}
\caption{Query for dangers possibilities on district \textbf{StareMiasto}}
\label{fig:query}
\end{figure}

We can take a closer look at another example of reasoning. This time we will ask the ontology to fetch dangers, which may occur on district named \textbf{StareMiasto}. There is no class prepared for answering such a question on class hierarchy shown in Figure \ref{fig:assertedClassHierarchy}. We have to write the query directly, using DL Query tab in Protégé. Below, in Figure \ref{fig:query}, the complete query is prepared. 

\medskip

\noindent More precisely, the query fetches all classes of dangers, raised as a results of road conditions appeared on district \textbf{StareMiasto}.

We have to inform ontology about some details, for associated reasoner to give the correct answer. I would like to show how it can be done using Protégé editor. It is why I have prepared simple tutorial. In the next few steps we will add required information. The information we are going to add is appropriate just for this example, and can be very different in reality.

\begin{framed}
\noindent The analogical process, of filling ontology with additional knowledge, is called \textit{synchronization} in this paper. It is done automatically, as a part of functionality provided by a system described in the Chapter \ref{cha:trafficDangerWebSystem}. The detailed information will be provided later.
\end{framed}

\noindent We are going to execute following steps using Protégé:

\begin{enumerate}
    \setlength{\itemsep}{0cm}
    \setlength{\parskip}{0cm}

    \item Create individuals: postal code \textbf{30-020}, street \textbf{Szpitalna}, and district \textbf{StareMiasto}. The postal code should belong to street, and the street should belong to district (through \textbf{hasLocation} relation). Filling ontology with new individuals data is done using Protégé \textit{Individuals} tab:
    \begin{itemize}
        \setlength{\itemsep}{0cm}
        \setlength{\parskip}{0cm}

        \item add \textbf{hasLocation} relation to appropriate individual, by repeating following process for postal code and street:
        \begin{itemize}
            \setlength{\itemsep}{0cm}
            \setlength{\parskip}{0cm}

            \item select appropriate individual on \textit{Individuals} tree,
            \item click on \textit{Object property assertions} button in the \textit{Property assertions} window,
            \item in the opened window write \textit{"hasLocation Szpitalna"} for postal code \textbf{30-020} and \textit{"hasLocation StareMiasto"} for \textbf{Szpitalna} street,
        \end{itemize}

        \item define types for provided individuals, by repeating following process for postal code, street and district:
        \begin{itemize}
            \setlength{\itemsep}{0cm}
            \setlength{\parskip}{0cm}

            \item select appropriate individual on \textit{Individuals} tree,
            \item click on \textit{Types} button on \textit{Description} tab,
            \item in the opened window write appropriate type for each individual: \textit{"PostalCodeLocation"} for \textbf{30-020}, \textit{"StreetLocation"} for \textbf{Szpitalna}, and \textit{"DistrictLocation"} for \textbf{StareMiasto}.
        \end{itemize}

    \end{itemize}

    \item Associate the location defined by postal code \textbf{30-020} with \textbf{HighCongestionCondition} condition, to assert that such a condition occurs in this area. That process is done using \textit{Classes} tab:
    \begin{itemize}
        \setlength{\itemsep}{0cm}
        \setlength{\parskip}{0cm}

        \item select \textit{HighCongestionCondition} node on \textit{Asserted class hierarchy} tree,
        \item click \textit{Superclasses} button,
        \item write \textit{"hasLocation value 30-020"} in the newly opened window.
    \end{itemize}

\end{enumerate}

\begin{figure}[htp]
\centering
\includegraphics[scale=0.7]{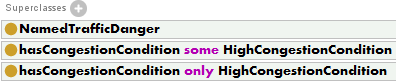}
\caption{\textbf{TrafficCongestionDanger} concept superclasses set}
\label{fig:trafficCongestionDanger}
\end{figure}

\noindent Now ontology is prepared. We can execute query shown in Figure \ref{fig:query} using DL Query tab. After executing, there will be one subclass in the result set. It is \textbf{TrafficCongestionDanger}. Take a closer look on description of that class, in order to understand why it has been fetched. In Figure \ref{fig:trafficCongestionDanger} we can see, that \textbf{TrafficCongestionDanger} class is connected to \textbf{HighCongestionDanger} condition.

\newpage

\begin{figure}[htp]
\centering
\includegraphics[scale=0.6]{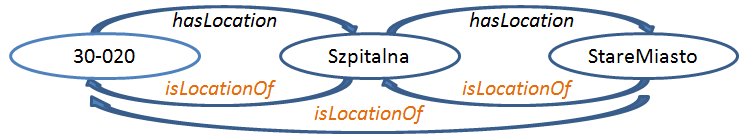}
\caption{Connections chain from postal code to district}
\label{fig:connection}
\end{figure}

\noindent Taking under consideration connections between postal code, street, and district individuals (Figure \ref{fig:connection}), reasoner can assume, that postal code area \textbf{30-020} belongs to \textbf{StareMiasto} district. 

\medskip

\begin{figure}[htp]
\centering
\includegraphics[scale=0.7]{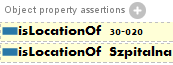}
\caption{Inferred property assertions for \textbf{StareMiasto} district}
\label{fig:stareMiasto}
\end{figure}

\noindent The reasoner deduction for \textbf{StareMiasto} is shown in Figure \ref{fig:stareMiasto}.

\medskip

\noindent In the above steps, we have synchronized the core ontology with sample knowledge. The way we have done that (information shown in Figure \ref{fig:trafficCongestionDanger} and Figure \ref{fig:stareMiasto}) allow to deduct, that \textbf{TrafficCongestionDanger} occurs on \textbf{StareMiasto} district.

\section{Summary}
\label{sub:ontologyDevelopementSummary}

\begin{figure}[htp]
\centering
\includegraphics[scale=0.7]{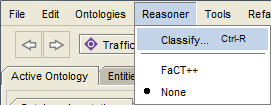}
\caption{Reasoner invoking in Protégé}
\label{fig:classify}
\end{figure}

Modern ontology development tools such as Protégé allow users to exploit ontologies conveniently, and provide intelligent guidance to find mistakes similar to a debugger in a programming environment. The only thing users have to do, for classes classification and inconsistencies detection process to be executed, is to choose a \textit{Classify} option shown in Figure \ref{fig:classify}.

\newpage

\noindent In addition, Protégé is ideal as a rapid prototyping environment in which ontology designers can instantly create individuals of specific classes in their ontology and experiment with semantic restrictions \cite{OntDrivDev}.

Furthermore, as a support for ontology researchers, Protégé has an open architecture and is easily extensible. That allows programmers to integrate arbitrary components with the tool. As a result ontology designers have a wide range of third-party plugins, which they can use. Protégé plugin library can be found under wiki page of the tool \cite{ProtegeWiki}. 

\chapter{Ontology-driven web system}
\label{cha:trafficDangerWebSystem}

\section{Introduction}
\label{sec:introductionTochapter4}

As it was mentioned at the beginning of the thesis, the system creation aim is an attempt to integrate database and ontology approach, for storing and inferring desired information about some domain in real time. In this chapter, such a composite approach is presented, in which location details of traffic conditions are stored in database, while clean abstract of traffic danger concept is described by core ontology. Such an integration results in dynamic deduction possibility of desired knowledge, using the ontology-based approach, instead of using static relations defined in database only.

The program cooperates with traffic danger ontology, and synchronizes that ontology with information stored in database. Next part of the process is traffic dangers inference. In this part reasoner searches for dangers, which occur on selected locations. Traffic dangers fetch the criteria only, when potentially dangerous traffic conditions are localized on area, we are looking for dangers. Locations of traffic conditions occurrences are defined by postal codes. In database, postal codes are connected with streets. Streets in turn are connected with districts. Ontology describes the way of connections between mentioned locations. Because of that fact, ontology gives the possibility of answering, which kind of traffic danger can occur on the desired location (postal code, street or district). As it was said before, the deduction is based on specific traffic conditions connected to specific postal codes.

Users are allowed to create dynamic questions, and getting results of inferred traffic dangers. That functionality is provided by the front dashboard page. Additionally, system allows trusted users to make changes to locations of traffic conditions occurrence. For working with latest data, provided by trusted users, synchronization mechanism is implemented. Synchronization integrates core ontology, describing the abstract of traffic dangers, with specific real time data. Synchronization process is executed first time on application startup. The startup can be understand as first request to the server while accessing main page. That functionality is also available on demand, after pressing synchronization button. After synchronization, we are cooperating with ontology cached in memory.

The project is coded in Java. Dependencies management and versioning is the task of Apache Maven tool. Main technologies used inside are: PostgreSQL, Hibernate, JSP, Spring MVC, jQuery, The OWL API, HermiT and Log4j.

\section{Ontology-driven software development}
\label{sec:ontologyDrivenSoftwareDevelopment}

An important aspect in developing any Semantic Web application, including application described in that chapter, is the ontology-driven architecture. In a simplification ontology-driven concept is extracted below:

\medskip

\begin{figure}[htp]
\centering
\includegraphics[scale=0.5]{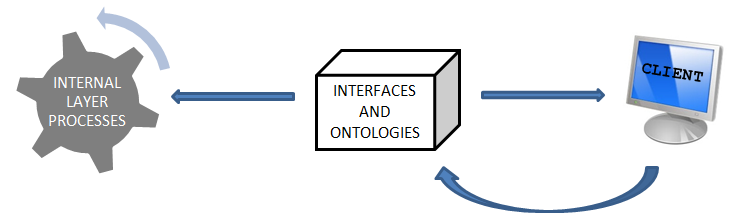}
\caption{Ontology-driven approach}
\label{fig:ontologyDrivenConcept}
\end{figure}

\noindent We can see that Semantic Web application in general is divided into 2 separate but linked layers \cite{OntDrivDev}:
\begin{itemize}
    \setlength{\itemsep}{0cm}
    \setlength{\parskip}{0cm}

    \item internal layer - contains all software logic: storage, control and deduction processes,
    \item Semantic Web layer - makes ontologies and interfaces available to the public.
\end{itemize}

\noindent Semantic Web layer is the external part. Interfaces of that layer are used to control the internal behavior, in particular the outcome of reasoning algorithms. Because of reusability of the ontologies, it is required to spend relatively lot of effort to provide good design and tests of that part. Domain experts should be involved in developing these tasks in parallel with programmers, to provide short iterations in developing as a results of quick responses for every inconsistencies in ontology like irrelevant hidden relationships, domain descriptions mistakes or usability problems. 

\subsection{Agile methodology}
\label{sub:agileMethodology}

All of that suggests, that software development based on agile methodologies is going to be the very dedicated while working on Semantic Web systems. Agile software development is based on iterative development, where requirements and solutions evolve through collaboration between self-organizing cross-functional teams. The term was introduced in the Agile Manifesto in 2001, which says:

\newpage

\noindent \textit{"We are uncovering better ways of developing software by doing it and helping others do it. Through this work we have come to value:}

\begin{itemize}
    \setlength{\itemsep}{0cm}
    \setlength{\parskip}{0cm}

    \item[] \textit{\textbf{individuals and interactions} over processes and tools,}
    \item[] \textit{\textbf{working software} over comprehensive documentation,}
    \item[] \textit{\textbf{customer collaboration} over contract negotiation,}
    \item[] \textit{\textbf{responding to change} over following a plan.}
\end{itemize}

\noindent \textit{That is, while there is value in the items on the right, we value the items on the left more"} \cite{AgileManifesto}.

\subsection{Model-driven architecture}
\label{sub:modelDrivenArchitecture}

Ontology-driven software design approach has a lot in common with model-driven architecture (MDA). MDA movement was launched by the Object Management Group (OMG) in 2001. That approach shows \textit{"how to better integrate high-level domain models into the development cycles of conventional software"} \cite{OntDrivDev}.

A central idea of MDA is to defines system functionality using a platform-independent model (PIM) using an appropriate domain-specific language (DSL). The PIM is next translated to one or more platform-specific models (PSMs) that computers can run. In the other words the goal of MDA is to employ languages like UML and generate from them a code appropriate for specific platform. Ontology-driven software development follows similar idea, but the way it does it is much more aggressive. It is because domain models can be used not only for code generation, but what is more remarkable, they can be involved in run time executable tasks. 

Nevertheless, any progress in MDA technology and tools may be useful for the Semantic Web community. In addition, MDA has aggregated wide range of modeling language standards under one metamodel writing standard called Meta-Object Facility (MOF). The OMG has created an initiative for rapid development ontology-related technologies. One of the OMG's efforts are directed towards mapping definition between OWL and MOF/UML. That integration can become a key for bringing that technologies much closer together. Because OWL does not need to be the best modeling language for everyone who is involved in domain design tasks, domain-specific languages can be used, depending on the task and
the expertise of the domain modelers. Platform dependent languages can be then translated into the details of OWL or any other desired language.

\newpage

\section{System use cases}
\label{sec:systemUseCases}

On the diagram below, actors and use cases of the system are shown:

\medskip

\begin{figure}[htp]
\centering
\includegraphics[scale=0.58]{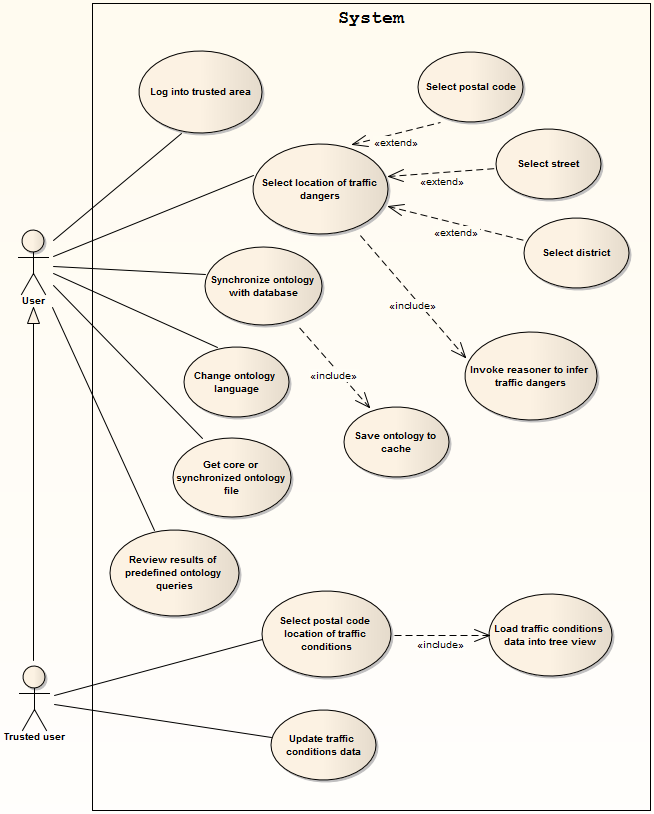}
\caption{Use cases of the system}
\label{fig:useCases}
\end{figure}

\newpage

\section{Interaction with system for updating and inferring data}
\label{sec:interaction}

On the diagram below, sequence diagram for updating and inferring data is shown:

\medskip

\begin{figure}[htp]
\centering
\includegraphics[scale=0.58]{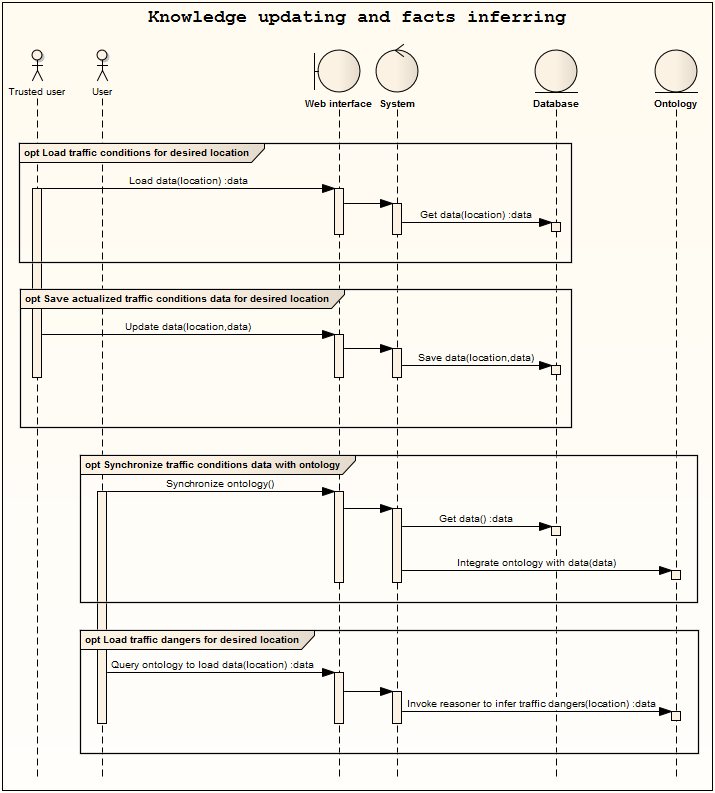}
\caption{Sequence diagram for updating and inferring data}
\label{fig:sequence}
\end{figure}

\newpage

\section{System flow}
\label{sec:systemFlow}

On the diagram below, data flow in the system is shown:

\medskip

\begin{figure}[htp]
\centering
\includegraphics[scale=0.5]{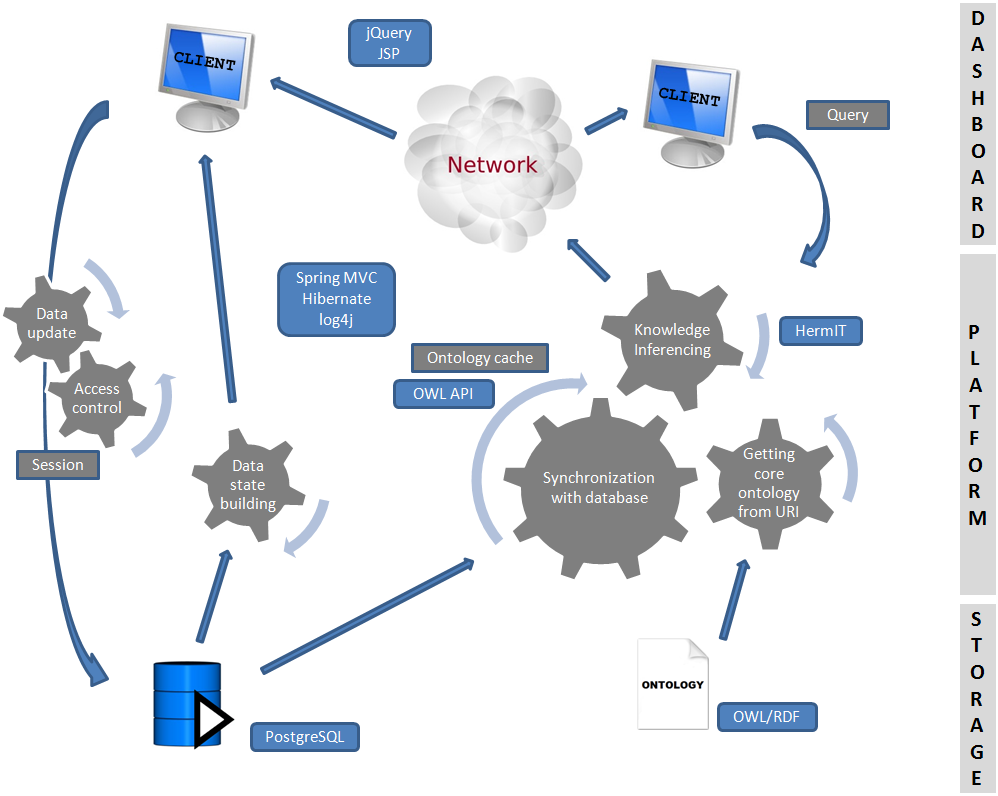}
\caption{Data flow in the system}
\label{fig:flow}
\end{figure}

\begin{tabular}[htb]{m{2cm}m{6cm}m{2cm}m{6cm}}
    \includegraphics[scale=0.7]{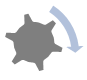}      & Operation         & \includegraphics[scale=0.7]{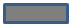}       & Annotation \\
    \includegraphics[scale=0.7]{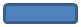}     & Technology used   & \includegraphics[scale=0.7]{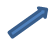}       & Internal system flow \\
    \includegraphics[scale=0.7]{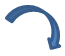}      & User input        & \includegraphics[scale=0.7]{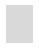}   & System layer
\end{tabular}

\newpage

\noindent The system is divided into 3 functionally different layers: web dashboard layer cooperating with users (through browser clients), platform layer which is the core of system and storage layer, where data is stored, including ontology which drives the system. 

On the left part of the diagram we can see session established for trusted users (after authentication pass), who can modify system database data. Database data can be modified through assigning specified information to dynamically built tree of current data state. 

On the top, diagram shows, that all users can interact with dashboard for querying system, to get desired information. 

In the center of the figure there are 3 processes: downloading of the core ontology, synchronization core ontology with current data uploaded by trusted users and inferring ontology dependencies. Core ontology can be stored on local or remote server and is accessed by URI of location. Synchronization is based on OWL API library, and provides fresh information (cached in memory) for semantic reasoner to infer, as a response to end users queries. Deduction of classes is provided by HermiT reasoner. 

Cooperation with database is provided through Hibernate ORM technology. User interface is built with Java Server Pages and jQuery JavaScript library, while requests from users and appropriate responses, are controlled by Spring MVC. For logging the results of particular operations, Log4j library is used. Ontology can be provided in different formats like OWL2 XML, RDF/XML or Manchester Syntax. PostgreSQL is chosen as SQL database server. All of the mentioned technologies are free and open source.

\begin{framed}
\noindent It is just a brief preview of the system main tasks and particular technologies. The layers cooperation will be explained in detail in the next chapters. There will be also very short preview of mentioned technologies, which are used to build the project.
\end{framed}

\newpage

\section{Database structure}
\label{sec:databaseStructure}

\begin{figure}[htp]
\centering
\includegraphics[scale=0.7]{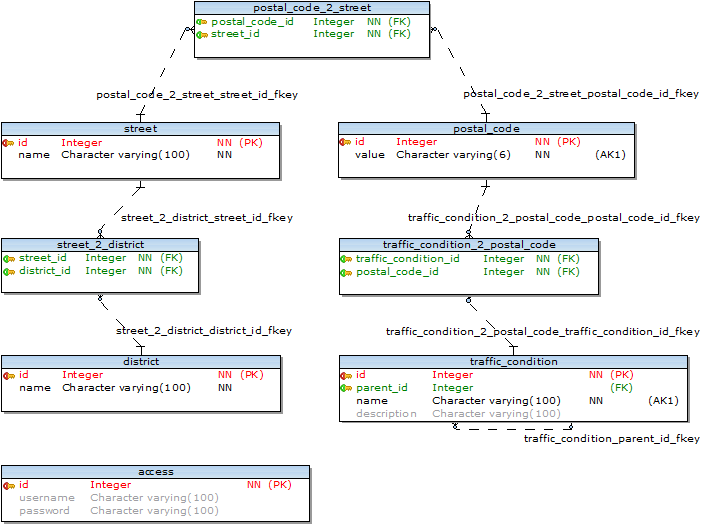}
\caption{ER diagram of \texttt{traffic} database}
\label{fig:erd}
\end{figure}

Entity relationship diagram is shown in Figure \ref{fig:erd}. The diagram was made using Toad Data Modeler \cite{CasestudioHome}.

\medskip

\noindent Database tables structure consist of tables describing locations: \texttt{street}, \texttt{district} and \texttt{postal\_code}, tables handling many-to-many relationships: \texttt{postal\_code\_2\_street}, \texttt{street\_2\_district}, \texttt{traffic\_condition\_2\_postal\_code}, table with traffic conditions structure \texttt{traffic\_condition} and table providing authentication for users: \texttt{access}.

\newpage

\begin{tabularx}{\textwidth}{X>{\hsize=10.5cm}X}
    \vtop{\vskip 0pt \vskip -\ht\strutbox 
    \begin{tabular}{|l|l|}
        \hline
        \multicolumn{2}{|c|}{\texttt{street}} \\
        \hline 
        \textcolor{blue}{\texttt{id}} & street unique identifier \\
        \textcolor{blue}{\texttt{name}} & street name \\
        \hline
    \end{tabular}
    \vskip -\dp\strutbox }%
    & Table contains streets names, used for defining locations of specific traffic conditions. \\  
\end{tabularx} 

\begin{tabularx}{\textwidth}{X>{\hsize=10.2cm}X}
    \vtop{\vskip 0pt \vskip -\ht\strutbox 
    \begin{tabular}{|l|l|}
        \hline
        \multicolumn{2}{|c|}{\texttt{district}} \\
        \hline
        \textcolor{blue}{\texttt{id}} & district unique identifier \\
        \textcolor{blue}{\texttt{name}} & district name \\
        \hline
    \end{tabular}
    \vskip -\dp\strutbox }%
    & Table contains district names, used for defining locations of specific traffic conditions. \\  
\end{tabularx}

\begin{tabularx}{\textwidth}{X>{\hsize=9.3cm}X}
    \vtop{\vskip 0pt \vskip -\ht\strutbox 
    \begin{tabular}{|l|l|}
        \hline
        \multicolumn{2}{|c|}{\texttt{postal\_code}} \\
        \hline
        \textcolor{blue}{\texttt{id}} & postal code unique identifier \\
        \textcolor{blue}{\texttt{value}} & postal code value \\
        \hline
    \end{tabular}
    \vskip -\dp\strutbox }%
    & Table contains postal codes values, used for defining locations of specific traffic conditions. \\  
\end{tabularx}

\begin{tabularx}{\textwidth}{X>{\hsize=6.2cm}X}
    \vtop{\vskip 0pt \vskip -\ht\strutbox 
    \begin{tabular}{|l|l|}
        \hline
        \multicolumn{2}{|c|}{\texttt{traffic\_condition}} \\
        \hline
        \textcolor{blue}{\texttt{id}} & traffic condition unique identifier \\
        \textcolor{blue}{\texttt{parent\_id}} & parent traffic condition unique identifier \\
        \textcolor{blue}{\texttt{name}} & traffic condition unique name \\
        \textcolor{blue}{\texttt{description}} & traffic condition description \\
        \hline
    \end{tabular}
    \vskip -\dp\strutbox }%
    & Table contains traffic conditions names and descriptions. Besides that table defines relations between traffic conditions structure. \\  
\end{tabularx}

\begin{tabularx}{\textwidth}{X>{\hsize=8.6cm}X}
    \vtop{\vskip 0pt \vskip -\ht\strutbox 
    \begin{tabular}{|l|l|}
        \hline
        \multicolumn{2}{|c|}{\texttt{street\_2\_district}} \\
        \hline
        \textcolor{blue}{\texttt{street\_id}} & street unique identifier \\
        \textcolor{blue}{\texttt{district\_id}} & district unique identifier \\
        \hline
    \end{tabular}
    \vskip -\dp\strutbox }%
    & Table maps streets to districts, because one street can belong to many districts as well as one district can gather many streets. \\  
\end{tabularx}

\begin{tabularx}{\textwidth}{X>{\hsize=7.3cm}X}
    \vtop{\vskip 0pt \vskip -\ht\strutbox 
    \begin{tabular}{|l|l|}
        \hline
        \multicolumn{2}{|c|}{\texttt{street\_2\_postal\_code}} \\
        \hline
        \textcolor{blue}{\texttt{street\_id}} & street unique identifier \\
        \textcolor{blue}{\texttt{postal\_code\_id}} & postal code unique identifier \\
        \hline
    \end{tabular}
    \vskip -\dp\strutbox }%
    & Table maps streets to postal codes, because one street can gather many postal codes as well as one postal code can belong to many streets. \\  
\end{tabularx}

\begin{tabularx}{\textwidth}{X>{\hsize=5.1cm}X}
    \vtop{\vskip 0pt \vskip -\ht\strutbox 
    \begin{tabular}{|l|l|}
        \hline
        \multicolumn{2}{|c|}{\texttt{traffic\_condition\_2\_postal\_code}} \\
        \hline
        \textcolor{blue}{\texttt{traffic\_condition\_id}} & traffic condition unique identifier \\
        \textcolor{blue}{\texttt{postal\_code\_id}} & postal code unique identifier \\
        \hline
    \end{tabular}
    \vskip -\dp\strutbox }%
    & Table maps traffic conditions to postal codes, because one traffic condition can occur on location defined by many postal codes as well as one postal code can define location of many traffic conditions occurred in parallel. \\  
\end{tabularx}

\begin{tabularx}{\textwidth}{X>{\hsize=10.8cm}X}
    \vtop{\vskip 0pt \vskip -\ht\strutbox 
    \begin{tabular}{|l|l|}
        \hline
        \multicolumn{2}{|c|}{\texttt{access}} \\
        \hline
        \textcolor{blue}{\texttt{id}} & access id \\
        \textcolor{blue}{\texttt{username}} & user login \\
        \textcolor{blue}{\texttt{password}} & user password \\
        \hline
    \end{tabular}
    \vskip -\dp\strutbox }%
    & Table contains credentials of trusted users. The password is encoded in database with SHA-1 function. \\  
\end{tabularx}

\section{Synchronization}
\label{sec:synchronization}

One of the most important aspects of the system is possibility of integration data from database with ontology data. I have called that process as \textbf{synchronization}. It allows to complete traffic danger ontology with additional information from database. These additional knowledge consist of locations structure and traffic conditions occurrence in that locations.

This approach provides loose coupling between core ontology, describing the abstract of traffic dangers, and synchronized ontology, which is filled with specific data connected with real time conditions on specific area. Generally, in computing, coupling refers to the degree of direct knowledge that one class has of another. Loose coupling was introduced by Karl Weick \cite{Wei76}. The term can also refer to our case. The only difference is that instead of classes we are now talking about ontologies. Core ontology describes clear concept of traffic danger, while synchronized one is related to specified environment. In the other words, synchronized ontology can differ on various environments where it is deployed. For example traffic conditions information for Krakow are very different than those for Warsaw. Of course we can synchronize our ontology at once with all global data, but it can result in system overloading and decreasing performance while inferring dependencies.

\section{System overview}
\label{sec:systemOverview}

\subsection{Traffic danger board}
\label{sub:trafficDangerBoard}

When user types URL of the system board page into the browser, e.g. \url{http://localhost:8080/traffic_web-1.0.0/board.html}, he will see the front page of the system. Overview of this page, called \textit{Traffic Danger Board}, is shown in Figure \ref{fig:boardOverall}.

\afterpage{%
\begin{landscape}
\centering\vspace*{\fill}
\begin{figure}[htp]
\centering
\includegraphics[scale=0.45]{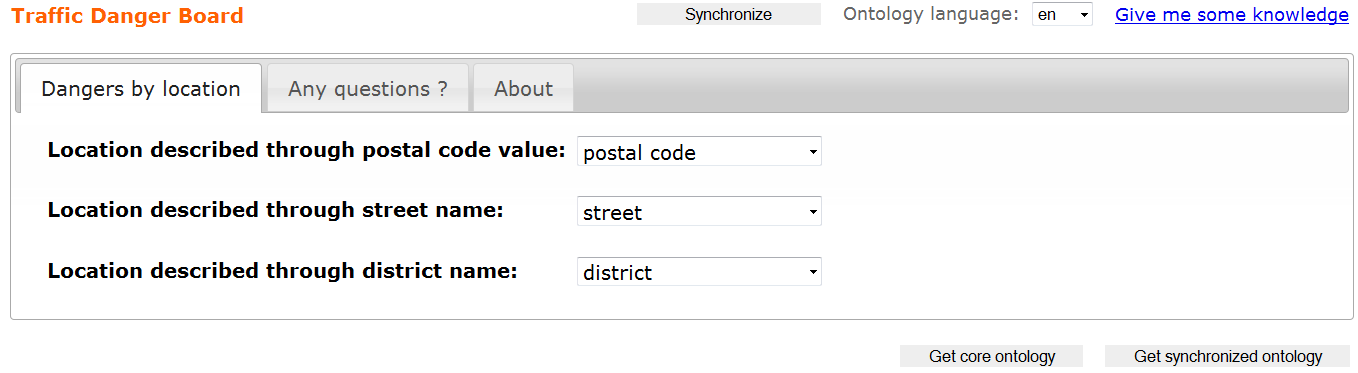}
\caption{Overview of the system front page}
\label{fig:boardOverall}
\end{figure}
\vfill
\end{landscape}
}

When this first request to the board page is called, the internal synchronization process is also executed. After the request is completed, we have the possibility of interaction with ontology which is integrated with current traffic conditions data. Such a synchronized ontology is cached in memory. It is why the execution time of first request is longer than further requests. Further requests, connected with refreshing page or invoking reasoner related operations, are very fast, because synchronization process is not being recalled. All system operations are processed on cached ontology. 

\newpage

\begin{figure}[htp]
\centering
\includegraphics[scale=0.6]{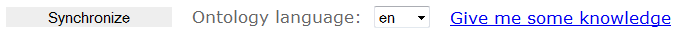}
\caption{Top toolbar of main page}
\label{fig:boardTopToolbar}
\end{figure}

\noindent If we want to synchronize our ontology with the latest data from database, we have to explicitly invoke synchronization task. It is provided after pushing button called \textit{Synchronize}. That button is located in top right corner of main page, in the toolbar. Top toolbar of the dashboard is shown in Figure \ref{fig:boardTopToolbar}.

\medskip

\begin{figure}[htp]
\centering
\includegraphics[scale=0.6]{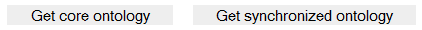}
\caption{Bottom toolbar of main page}
\label{fig:boardBottomToolbar}
\end{figure}

\noindent In the bottom toolbar (Figure \ref{fig:boardBottomToolbar}), we have the possibility of downloading 2 kinds of ontologies: original (core) and updated (synchronized). It is just a simple feature, which can be helpful for developers or other interested people who want to preview ontology in a raw state - from file, or by loading such a file into an ontology editor such as Protégé.

\medskip

\begin{figure}[htp]
\centering
\includegraphics[scale=0.55]{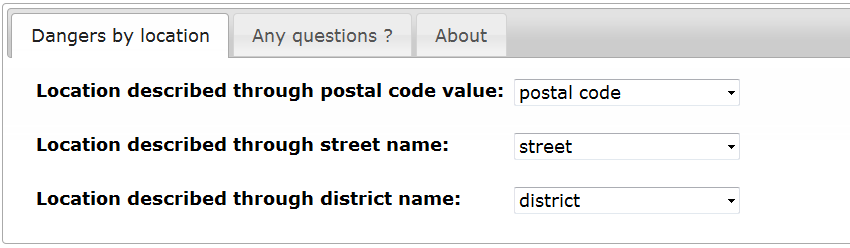}
\caption{Dangers by location}
\label{fig:boardMain}
\end{figure}

\noindent Locations data in database are well known. That kind of information is fixed and depends on area we are working with. Based on provided postal codes, we can assign traffic conditions to selected ones (which are further connected to streets and districts). Traffic conditions structure in database is identical as traffic condition subtree structure defined in ontology, which was previously shown in Figure \ref{fig:assertedClassHierarchy}. In Figure \ref{fig:boardMain} we can see core part of the main page.

\medskip

\begin{figure}[htp]
\centering
\includegraphics[scale=0.5]{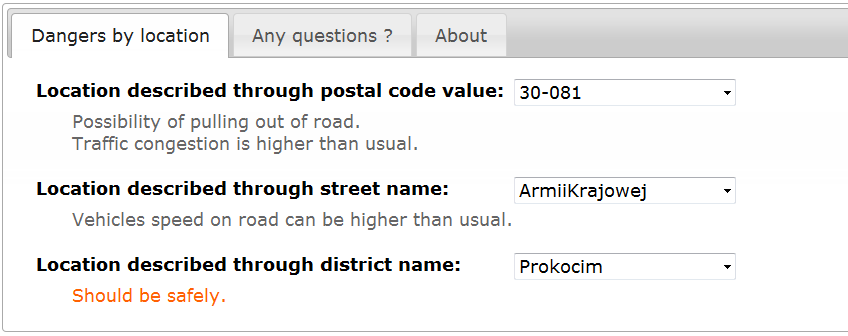}
\caption{Dangers by location - inferred values}
\label{fig:boardMainInferred}
\end{figure}

\noindent The main page is divided into 3 tabs. First one, named \textit{Dangers by location}, allows for previewing dangers on desired locations. We can choose dangers connected directly to desired postal code, or indirectly to street or even wider - to district. After choosing desired location, system queries ontology for occurred traffic dangers. Ontology querying means here that system invokes reasoner on cached ontology, and demands answer for question similar to that shown in Section \ref{sub:secondExample}. This time, in the place of \textbf{StareMiasto}, there can be inserted any selected location value. It is shown in Figure \ref{fig:boardMainInferred}.

\newpage

\noindent The second tab (called \textit{Any questions?}), shown in Figure \ref{fig:boardQuestions}, displays responses for example predefined questions, our ontology is able to answer.

\afterpage{%
\begin{landscape}
\centering\vspace*{\fill}
\begin{figure}[htp]
\centering
\includegraphics[scale=0.45]{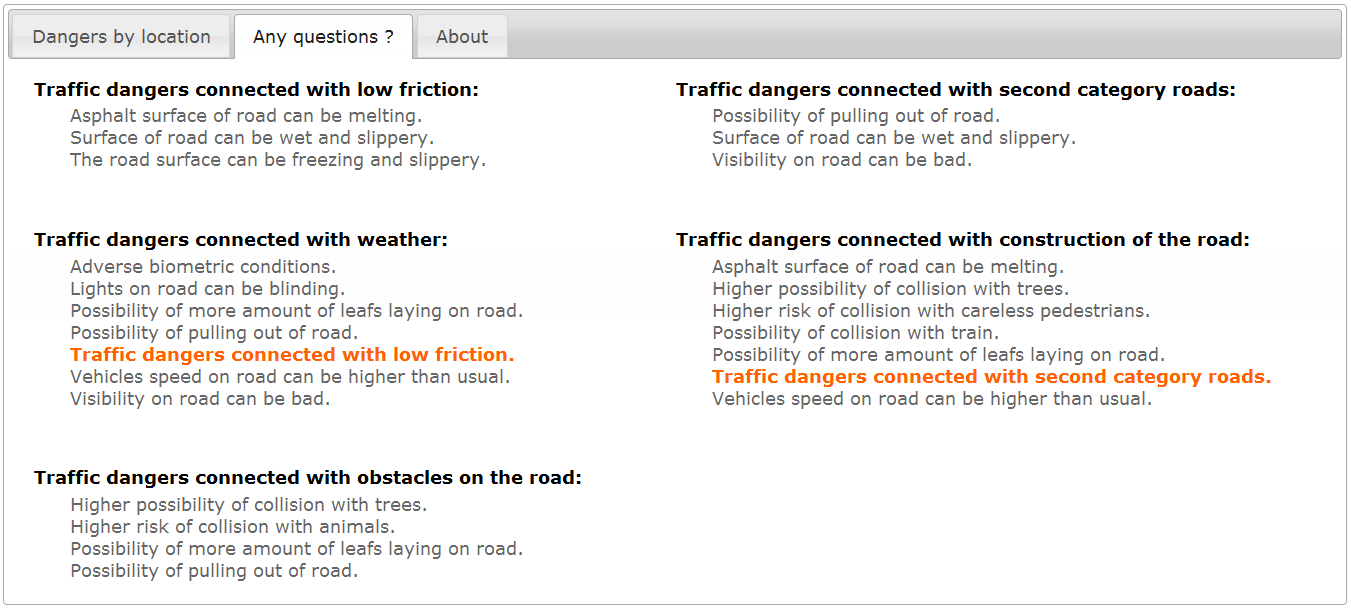}
\caption{Ontology responses for predefined questions}
\label{fig:boardQuestions}
\end{figure}
\vfill
\end{landscape}
}

These questions are not build dynamically on runtime, just like it was always after changing drop down lists values on the previous tab. They are defined statically in the ontology, as defined classes. Defined classes topic was explained earlier in Chapter \ref{cha:trafficDangerOntology}. In the practical shortcut, the reasoner can only automatically classify classes under classes which are defined, because they have fulfilled all necessary and sufficient conditions for reasoning to happen. From the technical point of view, to get answers for such questions, the system has to invoke reasoner on ontology. After that operation, dependencies of all defined classes will be automatically deducted. Developer has to get all that inferred dependent subclasses and provide them to the user. In OWLAPI, the OWL parsing library, that is much easier task to do, than building complex questions dynamically. 

\begin{framed}
\noindent Working with OWLAPI while doing more complex operations, can be a bit confusing for developers, who have never been working with ontologies before. Nevertheless, after review of source code examples provided by the author, library usage will be straightforward. These examples can be found inside trunk branch of the library source code.
\end{framed}

\noindent The last tab on the page, called \textit{About}, provides a brief explanation of the system aim.

\bigskip

\noindent Dashboard provides also the possibility of changing language in which ontology displays data. Traffic danger ontology defines 2 languages: English and Polish. That is the reason why we have only that 2 options provided in drop down list shown on top toolbar (Figure \ref{fig:boardTopToolbar}). Various languages support is provided by the ontology itself. System will data in user preferred language, while fetching data from ontology. Database does not interfere with translations.

When the user choose \textit{Give me some knowledge} hyperlink, located in the right top corner of the dashboard (Figure \ref{fig:boardTopToolbar}), he will be redirected to login page, described in the next section.

\newpage

\subsection{Trusted area authentication form}
\label{sub:trustedAreaAuthenticationForm}

\begin{figure}[htp]
\centering
\includegraphics[scale=0.6]{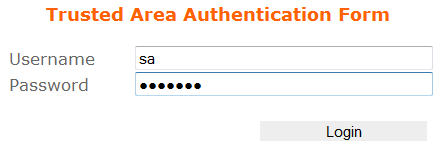}
\caption{Logging panel for trusted area}
\label{fig:trustedAreaAuthenticationForm}
\end{figure}

This page provides logging panel (Figure \ref{fig:trustedAreaAuthenticationForm}) for authenticate trusted users, who can modify the database state. 

\medskip

\noindent Database data modification is directly connected with results, users will get while working with ontology, just after synchronization with that updated data. That is the reason of restricted access to control panel. That is be strongly unexpected situation for users, to get inappropriate or even adulterated information about traffic dangers on dedicated areas. This can even provide to potentially unsafe situations on roads, as a result of faulty information about real traffic dangers possibilities. The access is widely restricted but provided for those, who declared inclination for filling database with correct and coherent data.

The database created from scripts, contains example trusted user - username: \texttt{sa}, password: \texttt{traffic}. The user is created only for presentation purposes and should be deleted in production environment. When user provides invalid credentials, appropriate message will be shown on the page.

After login, user will be persisted in the server session. The next entrance to \textit{Traffic Danger Control Panel} will be transparent for user, without necessity of redundant authentication.

\newpage

\subsection{Traffic danger control panel}
\label{sub:trafficDangerControlPanel}

This page allows trusted users for filling current information about specific traffic conditions locations. In the top left corner system displays user, who is actually logged in. The artificially provided user for presentations purposes is called \texttt{sa}:

\medskip

\begin{figure}[htp]
\centering
\includegraphics[scale=0.6]{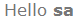}
\caption{Logged user name}
\label{fig:panelTopRight}
\end{figure}

\noindent Below we can see the base view of the page content:

\medskip

\begin{figure}[htp]
\centering
\includegraphics[scale=0.6]{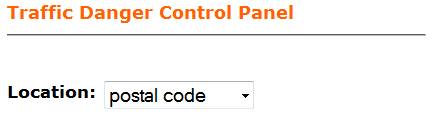}
\caption{Control panel raw view}
\label{fig:panelTopLeft}
\end{figure}

\begin{figure}[htp]
\centering
\includegraphics[scale=0.55]{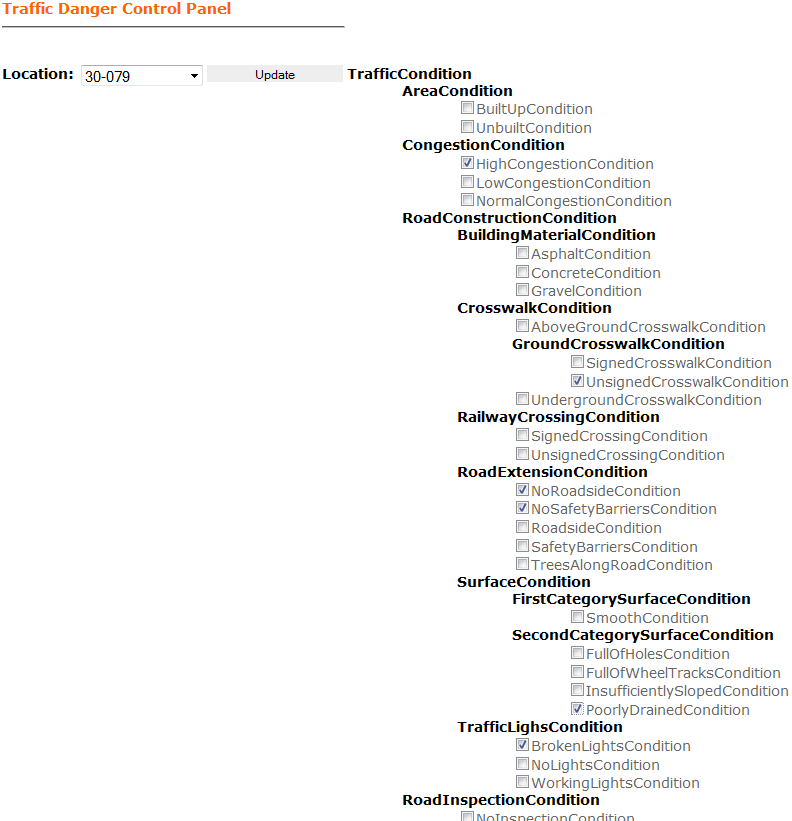}
\caption{Part of dynamically created view}
\label{fig:panelMain}
\end{figure}

\noindent After selection of desired postal code value from drop down list, system is building a dynamic tree of traffic conditions. The tree construction is based on database data. Traffic conditions structure in database (provided through table called \texttt{traffic\_condition} shown in Figure \ref{fig:erd} diagram) is identical, as traffic conditions tree structure in ontology, shown in Chapter \ref{cha:trafficDangerOntology} in Figure \ref{fig:assertedClassHierarchy}. ERD diagram of database schema shows, that \texttt{traffic\_conditions} table is connected with \texttt{postal\_code} through intermediate many-to-many relationship table \texttt{traffic\_condition\_2\_postal\_code}. Drop down list is filled with postal codes from database. After selection of desired postal code from the list, system is building current data structure. It is a traffic conditions tree, shown in Figure \ref{fig:panelMain}.

\newpage

\noindent Users can assign selected location to desired condition by checking radio button next to condition name. When the information is correct, it is ready to be persisted inside database. That is done by pushing \textit{Update} button, next to the drop down list. Knowledge is now updated and can be used for dangers deduction process on front dashboard page, just after the synchronization is done.

\newpage

\section{Technologies inside}
\label{sec:technologiesInside}

\subsection{Introduction}
\label{sub:introductionToTechnologies}

This chapter provides a short preview of main technologies used to build described system. The project is written in Java language in using of Eclipse Java EE IDE. Dependencies management and versioning is the task of Apache Maven tool \cite{MavenHome}.

\subsection{Brief overview}
\label{sub:briefOverview}

\subsubsection{PostgreSQL}
\label{sss:postgreSQL}

Database was designed using latest version of PostgreSQL, which can be downloaded from its home page \cite{PostgresHome}. According to documentation, \textit{"PostgreSQL is an object-relational database management system (ORDBMS) based on  POSTGRES, Version 4.2, developed at the University of California at Berkeley Computer Science Department. POSTGRES pioneered many concepts that only became available in some commercial database systems much later"} \cite{PostgresDocs}.

PostgreSQL is released under the PostgreSQL License, a liberal Open Source license, similar to the BSD or MIT licenses.

\subsubsection{Hibernate}
\label{sss:hibernate}

For interaction with database, system uses object-relational mapping (ORM) library called Hibernate. Hibernate is a library designed for Java language, and provides a framework for mapping an object-oriented domain model to a traditional relational database. Hibernate home page \cite{HibernateHome} provides all necessary information about that technology. There is also a great tutorial \cite{HibernateDocs}, just excellent for learning how to make not only the first steps, but also advanced operations.

One of the most primary features of Hibernate technology, is the possibility of mapping from Java classes to database tables (and from Java data types to SQL data types). Hibernate also provides data query and retrieval facilities. Hibernate automatically generates the SQL calls and relieves the developer from manual result set handling and object conversion, keeping the application portable to all supported SQL databases, with database portability delivered at very little performance overhead.

Hibernate is free software that is distributed under the GNU Lesser General Public License.

\subsubsection{Java Server Pages}
\label{sss:jsp}

According to home page of JavaServer Pages, \textit{"JSP technology enables Web developers and designers to rapidly develop and easily maintain, information-rich, dynamic Web pages that leverage existing business systems. As part of the Java technology family, JSP technology enables rapid development of web-based applications that are platform independent. JSP technology separates the user interface from content generation, enabling designers to change the overall page layout without altering the underlying dynamic content"} \cite{JSPHome}.

JSP technology allows developers to build pages using XML-like tags. Such tags encapsulates the logic that generates the content for the page. The application logic can reside in server-based resources (such as JavaBeans component architecture) that the page accesses with these tags. By separating the page logic from its design and supporting a reusable component-based design, JSP technology to fast and easy build web-based applications. Besides standard markup tags (HTML and XML), JSP uses also scriplet tags for building pages. Scriptlet elements are delimited blocks of Java code which may be intermixed with the markup tags.

JavaServer Pages technology is an extension of the Java Servlet technology. A servlet is a Java class which conforms to the Java Servlet API, a protocol by which a Java class may respond to HTTP requests. Therefore developers may use a servlet to add dynamic content to a web server using the Java platform. JSPs are compiled into servlets by a JSP compiler. The compiler can generate a servlet in Java code that is then compiled by the Java compiler. It can also compile the servlet to byte code which is directly executable. The bytecode must be executed within a Java virtual machine (JVM). It provides an abstract platform-neutral runtime environment. To sum up, servlets are platform-independent, server-side modules extending capabilities of a web server. 

\subsubsection{Spring MVC}
\label{sss:springMVC}

Model View Controller (MVC) is a software architecture which has its roots in Smalltalk \cite{Ree78}. Currently is considered an architectural pattern used in software engineering. 

Applications can contain a mixture of data access code, business logic code, and presentation code. Such applications are difficult to maintain, because interdependencies between all of the components cause strong ripple effects whenever a change is made anywhere. That situation is called as high coupling. Classes depends on so many other classes, that reusing is impossible. Adding new data access code or new data views often requires reimplementing or copping blocks of business logic code, which then requires maintenance in multiple places. The Model View Controller design pattern solves these problems by decoupling data access, business logic, and data presentation and user interaction \cite{SpringMVCSunBlueprints}. The following diagram represents the MVC pattern:

\newpage

\begin{figure}[htp]
\centering
\includegraphics[scale=0.7]{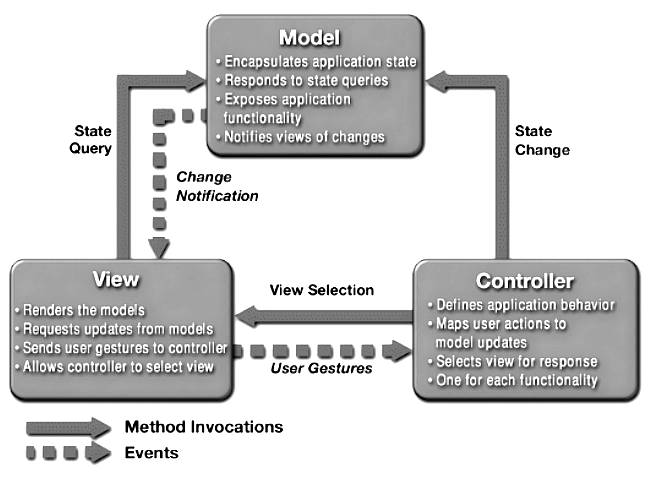}
\caption{Model View Controller \cite{SpringMVCSunBlueprints}}
\label{fig:mvc}
\end{figure}

\noindent SpringMVC in one of the implementations of model view controller pattern. It is a part of Spring Framework, an open source application framework for the Java platform. A key design principle in Spring Web MVC and in Spring in general is the \textit{open for extension, closed for modification} principle (OCP principle). Open for extension means, that the behavior of the module can be extended. The module behavior can be changed in new and various ways as the requirements of the application change, or to meet the needs of new applications. Closed for modification means that source code of module designed in this way is inviolate. There is not allowed to make source code changes to it \cite{Mar08}. 

I have used the latest version of SpringMVC, which has a lot of improvements to the previous versions. Simplification in development, part of dynamically evolving Spring Framework application platform and integration with JSP made me choose that technology to build the traffic danger web system. Good source of information about that technology can be found in the documentation on the home page of Spring Framework \cite{SpringHome}.

\subsubsection{jQuery}
\label{sss:jQuery}

According to jQuery home page, \textit{"jQuery is a fast and concise JavaScript Library that simplifies HTML document traversing, event handling, animating, and Ajax interactions for rapid web development. jQuery is designed to change the way that you write JavaScript"} \cite{jQueryHome}.

\newpage

jQuery is a cross-browser JavaScript library designed to simplify the client-side scripting. Cross-browsing refers to the ability for a websites or client-side scripts to support all the web browsers. jQuery's syntax simplifies navigation through the HTML documents, selection of DOM elements, handling events, creating animations and even building Ajax oriented applications in a gentle way. jQuery was released in January 2006 at BarCamp NYC by John Resig. Nowadays the library is used by over 27\% of the 10,000 most visited websites, jQuery is the most popular JavaScript library in use today \cite{jQueryUsageStatistics, UsageOfJavaScriptForWebsites}.

jQuery is free, open source software, under the terms of either the MIT License or the GNU General Public License (GPL) Version 2.

\subsubsection{The OWL API}
\label{sss:theOwlApi}

According to OWL APL home page, \textit{"the OWL API is a Java API and reference implementation for creating, manipulating and serializing OWL Ontologies. The latest version of the API is focused towards OWL 2"} \cite{OWLAPIHome}.

\bigskip

\noindent The OWL API includes the following components:
\begin{itemize}
    \setlength{\itemsep}{0cm}
    \setlength{\parskip}{0cm}

    \item an API for OWL 2 and an efficient in-memory reference implementation,
    \item RDF/XML parser and writer,
    \item OWL/XML parser and writer,
    \item OWL Functional Syntax parser and writer,
    \item Turtle parser and writer,
    \item KRSS parser,
    \item OBO Flat file format parser,
    \item reasoner interfaces for working with reasoners such as FaCT++, HermiT, Pellet and RACER.
\end{itemize}

\noindent The API is closely aligned with the OWL 2 structural specification. It supports parsing and rendering in the syntaxes defined in the W3C specification, namely, the Functional Syntax, RDF/XML, OWL/XML and the Manchester OWL Syntax. Library is written in Java. \textit{"The latest version of the OWL API has been designed to meet the needs of people developing OWL-based applications, OWL editors and OWL reasoners. It is a high level API that is closely aligned with the OWL 2 specification. It includes first class change support, general purpose reasoner interfaces, validators for the various OWL 2 profiles, and support for parsing and serializing ontologies in a variety of syntaxes. The API also has a very flexible design that allows third parties to provide alternative implementations for all major components"} \cite{Hor09}.

The OWL API is open source and is available under the LGPL License.

\subsubsection{HermiT}
\label{sss:hermiT}

According to HermiT home page, \textit{"HermiT is reasoner for ontologies written using the Web Ontology Language (OWL). Given an OWL file, HermiT can determine whether or not the ontology is consistent, identify subsumption relationships between classes, and much more."}

\textit{"HermiT is the first publicly-available OWL reasoner based on a novel "hypertableau" calculus which provides much more efficient reasoning than any previously-known algorithm. Ontologies which previously required minutes or hours to classify can often by classified in seconds by HermiT, and HermiT is the first reasoner able to classify a number of ontologies which had previously proven too complex for any available system to handle. HermiT uses direct semantics and passes all OWL 2 conformance tests for direct semantics reasoners"} \cite{HermiTHome}.

The version (v1.2.3) used in the Traffic Danger Web System is fully compatible with OWLAPI 3.0.0. It is the reason I have chosen that reasoner for working with, because for the time of writing that thesis the Fact++ and Pellet reasoners was not compatible with the newest version of the OWLAPI.

HermiT is open-source and released under LGPL.

\subsubsection{Log4j}
\label{sss:log4j}

Apache Log4j is a Java-based logging utility. It was originally written by Ceki Gülcü and is now a project of the Apache Software Foundation.

\section{Possible directions of development}
\label{sub:possibleDirectionsOfDevelopment}

Future directions of project development can be focused on extensions such as Web Services. Web Services will give the possibility of communication with external systems. That external systems can be perceived as software agents. Their tasks could be focused on periodic connections to our primary system, getting latest information set (serialized into ontology), and creating statistics about traffic dangers. Statistics can visualize frequencies of desired dangers on specific area, classification of safety in desired district at the turn of fixed time, etc.

\section{Summary}
\label{sub:applicationDevelopementSummary}

To summarize, the development of Semantic Web applications with the support of great tool like Protégé, supported by DL reasoner, is quite inspiring development task. Such development process is called as \textit{driven by ontology}. With the support of agile methodologies (like extreme programming) where frequent feedback is crucial for fast and stable system implementation, such development process comes to be powerful. Ontologies can be developed directly by domain experts. Because of direct access to executable systems, charged by recently developed domain models, feedbacks can be made frequently. Best practices from agile methodologies, effective at delivering a particular outcome, can be used in development high-quality domain models. Domain experts may work together with programmers, which can guarantee coherent testing and fast implementation processes.

\chapter{Conclusion}
\label{cha:conclusion}

\textit{"The goal of Semantic Web research is to transform the Web from a linked document repository into a distributed knowledge base and application platform, thus allowing the vast range of available information and services to be more effectively exploited."}
\begin{flushright} Ian Horrocks \cite{HorSSF} \end{flushright}

\noindent According to Handbook of Knowledge Representation \cite{HLP08}, \textit{"OWL has become the most used KR language in the history of the field, not because of its particular representational power, but rather because it was designed to be a common syntax usable by many KR systems, to be webized for easier sharing of ontologies and concepts, and to be expressive enough for many problems without totally sacrificing scalability"}.

Increasingly number of tools designed for OWL is both cause and motivation for the community, to develop ontologies in many various fields, not only in the scope of Semantic Web. There are areas like: biology, medicine, geography, geology, astronomy, agriculture or defense, in which ontologies are becoming adopted \cite{HLP08}.

Modern ontology development tools such as Protégé allow users to maintenance and built ontologies in quick, efficient way. They provide intelligent guidance to find mistakes similar to a debugger in a programming environment because of description logics reasoners support. That is also the reason, why Protégé is ideal as a rapid prototyping environment in which ontology designers can instantly create individuals of specific classes in their ontology and experiment with semantic restrictions \cite{OntDrivDev}.

Development of Semantic Web applications with the support of recent powerful tools is quite inspiring development task. Ontology-driven development with the support of agile methodologies is very efficient implementation process. Ontologies can be developed directly by domain experts. Because of direct access to executable systems, charged by recently developed domain models, feedbacks can be made frequently. Best practices from agile methodologies, effective at delivering a particular outcome, can be used in development high-quality domain models. Domain experts may work together with programmers, which can guarantee coherent testing and fast implementation processes.

\appendix
\setlength{\parindent}{0in}	

\chapter{Traffic danger web system deployment}
\label{cha:systemDeployment}

\section{Environment setup}
\label{sec:environmentSetup}

Java applications are typically compiled to bytecode that can run on any Java Virtual Machine (JVM) regardless of computer architecture. Because of Java platform independence, the system can be deployed on various environments.

\bigskip

For the project to be deployed, following applications should be pre-installed:
\begin{itemize}
    \setlength{\itemsep}{0cm}
    \setlength{\parskip}{0cm}

    \item PostgreSQL database engine,
    \item Java Runtime Environment (JRE),
    \item Apache Tomcat servlet container.
\end{itemize}

\subsection{PostgreSQL installation}
\label{sub:postgreSQLInstallation}

{\scshape Windows}
\smallskip

Get the latest version of PostgreSQL executable package from \url{http://www.postgresql.org/download/windows/}. After downloading simply double click on that package. The installer will do the job.

\bigskip

{\scshape Linux}
\smallskip

Get the latest version of PostgreSQL sources. For the date of writing that file latest version of PostgreSQL was v9.0beta2. Sources can be obtained by anonymous FTP from \url{ftp://ftp.postgresql.org/pub/source/v9.0beta2/postgresql-9.0beta2.tar.gz}. After downloading, unpack the file:
\begin{verbatim}
gzip -dc postgresql-9.0beta2.tar.gz | tar xf -
\end{verbatim}

This will create a directory \texttt{postgresql-9.0beta2} under the current directory with the PostgreSQL sources. Enter into that directory and execute installation procedure:

\newpage

\begin{verbatim}
./configure
gmake
su
gmake install
adduser postgres
mkdir /usr/local/pgsql/data/
chown postgres /usr/local/pgsql/data/
su - postgres
/usr/local/pgsql/bin/initdb -D /usr/local/pgsql/data/
/usr/local/pgsql/bin/postgres -D /usr/local/pgsql/data/ >logfile 2>&1 &
/usr/local/pgsql/bin/createdb test
/usr/local/pgsql/bin/psql test
\end{verbatim}

\subsection{Java Runtime Environment (JRE) installation}
\label{sub:jreInstallation}

Download the Java 2 Standard Edition Runtime (JRE) release version 5.0 or later, from \url{http://www.java.com/en/download/manual.jsp} and install the JRE according to the instructions included with the release. 

\bigskip

Set the environment variable named \texttt{JRE\_HOME} to the pathname of the directory into which you installed the JRE, e.g.

\bigskip

{\scshape Linux}
\begin{verbatim}
# for Bourne, bash, and related shells
export JRE_HOME=/usr/local/java/jre5.0
\end{verbatim}
\begin{verbatim}
# for csh and related shells
setenv JRE_HOME=/usr/local/java/jre5.0
\end{verbatim}

{\scshape Windows}
\begin{verbatim}
set JRE_HOME=C:\jre5.0
\end{verbatim}

You can also use the full JDK rather than just the JRE. In this case set the \texttt{JAVA\_HOME} environment variable to the pathname of the directory into which you installed the JDK.

\subsection{Apache Tomcat installation}
\label{sub:apacheTomcatInstallation}

Download the latest version of Tomcat binary distribution from \url{http://tomcat.apache.org/}, appropriate for your system and unpack the file into convenient location so that the distribution resides in its own directory:

\bigskip

{\scshape Linux}
\begin{verbatim}
cp apache-tomcat-6.0.26.tar.gz /usr/local/apache/
cd /usr/local/apache/
gzip -dc apache-tomcat-6.0.26.tar.gz | tar xf -
\end{verbatim}

\bigskip

Set the \texttt{CATALINA\_HOME} environmental variable (it is used to refer to the full pathname of the release directory):

\bigskip

{\scshape Linux}
\begin{verbatim}
# for Bourne, bash, and related shells
export CATALINA_HOME=/usr/local/apache/apache-tomcat-6.0.26
\end{verbatim}
\begin{verbatim}
# for csh and related shells
setenv CATALINA_HOME=/usr/local/apache/apache-tomcat-6.0.26
\end{verbatim}

{\scshape Windows}
\begin{verbatim}
set CATALINA_HOME=C:\apache\apache-tomcat-6.0.26
\end{verbatim}

\bigskip

{\large\textbf{Startup Tomcat:}}

\bigskip

{\scshape Linux}
\begin{verbatim}
$CATALINA_HOME/bin/startup.sh
\end{verbatim}

{\scshape Windows}
\begin{verbatim}
%CATALINA_HOME%\bin\startup.bat
\end{verbatim}

After starting the default web applications included with Tomcat will be available at address: \url{http://localhost:8080/}

\bigskip

\begin{framed}
If Tomcat is not responding, the reason can be another web server (or process) running and using provided port \texttt{8080}. The solution is to edit \path{$CATALINA_HOME/conf/server.xml} configuration file and change the default port.
\end{framed}

\bigskip

{\large\textbf{Shutdown Tomcat:}}

\bigskip

{\scshape Linux}
\begin{verbatim}
$CATALINA_HOME/bin/shutdown.sh
\end{verbatim}

{\scshape Windows}
\begin{verbatim}
%CATALINA_HOME%\bin\shutdown.bat
\end{verbatim}

\newpage

\section{Loading and configuration}
\label{sec:loadingAndConfiguration}

Open the \textit{Tomcat Manager} available under the \textit{Tomcat Administration} panel at address \url{http://localhost:8080/manager/html/}:

\begin{framed}
If you are not authorized to view this page, you should probably examine \path{$CATALINA_HOME/conf/tomcat_users.xml} file and, if necessary, define a new user with appropriate rights, e.g. \texttt{<user username="root" password="toor" roles="manager-gui" />}.
\end{framed}

\begin{figure}[htp]
\centering
\includegraphics[scale=0.6]{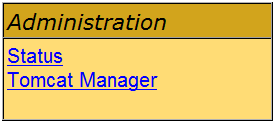}
\caption{Tomcat administration panel}
\label{fig:tomcatAdministrationPanel}
\end{figure}

\begin{figure}[htp]
\centering
\includegraphics[scale=0.5]{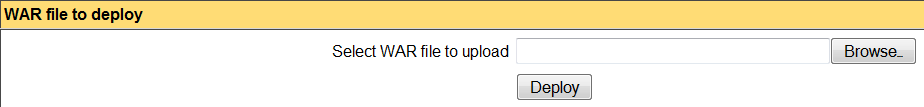}
\caption{WAR archive deployment section}
\label{fig:WARArchiveDeployment}
\end{figure}

Under the \textit{WAR file to deploy} section shown in Figure \ref{fig:WARArchiveDeployment}, deploy \texttt{traffic\_web-1.0.0.war} archive, containing prebuild web application for traffic danger web system.

\medskip

\bigskip

After deployment is complete, open for edition \path{$CATALINA_HOME/webapps/traffic_web/WEB-INF/web.xml} file and change value of \texttt{OntologyURI} parameter for appropriate path indicating ontology file for traffic danger \texttt{TrafficDanger.owl}. 

\bigskip

The URI can identify local or remote resource, for example:
\begin{itemize}
    \setlength{\itemsep}{0cm}
    \setlength{\parskip}{0cm}

    \item \path{file:///C:/Users/jwa/masters/TrafficDanger.owl} (local Windows location),
    \item \path{file:///home/jwa/masters/TrafficDanger.owl} (local Linux location),
    \item \path{http://host/TrafficDanger.owl} (remote location).
\end{itemize}

\smallskip

Final step is about SQL scripts execution, required for database creation. Run the following scripts in the given order: \texttt{postgres\_traffic\_user.sql}, \texttt{postgres\_traffic\_database.sql}, \texttt{postgres\_traffic\_schema.sql} and finally \texttt{postgres\_traffic\_data.sql}. The last one actually contains sample data, so its execution is optional.

\newpage

{\scshape Linux}
\smallskip

Change user to \texttt{postgres}, and change catalog to directory containing database scripts and invoke following commands:

\begin{verbatim}
psql -f postgres_traffic_user.sql
psql -f postgres_traffic_database.sql
psql -f postgres_traffic_schema.sql traffic
psql -f postgres_traffic_data.sql traffic
\end{verbatim}

After that, system should be ready to cooperate under address \url{http://localhost:8080/traffic_web-1.0.0/board.html}.

\setlength{\parindent}{0in}	

\chapter{Additional tools installation instructions}
\label{cha:additionalToolsInstallationInstructions}

\section{Protégé installation}
\label{sec:protegeInstallation}

\subsection{Editor installation}
\label{sub:editorInstallation}

The installation instructions can be found on Protégé home page \cite{ProtegeHome}. First, download the latest version of Protégé tool from the Protégé home page. Choose installation variant based on your system type.

\bigskip

{\scshape Windows}
\smallskip

After downloading double-click \texttt{install\_protege\_4.0.2.exe}.
\smallskip

\begin{framed}
If you do not have a Java virtual machine installed, be sure to download the package above which includes one. 
\end{framed}

\bigskip

{\scshape Linux}
\smallskip

After downloading open a shell and go to the directory of downloaded installer. At the prompt type:  
\begin{verbatim}
sh ./install_protege_4.0.2.bin 
\end{verbatim}

\begin{framed}
If you do not have a Java virtual machine installed, be sure to download the package above which includes one. Otherwise you may need to download one from Sun's Java website \cite{SunHome} or contact your OS manufacturer. 
\end{framed}

\newpage
{\scshape All other platforms}

\begin{enumerate}
    \setlength{\itemsep}{0cm}
    \setlength{\parskip}{0cm}

    \item Instructions for Unix or Unix-like operating systems
    \begin{itemize}
        \setlength{\itemsep}{0cm}
        \setlength{\parskip}{0cm}

        \item For Java 2, after downloading, type
\begin{verbatim}
java -jar install_protege_4.0.2.jar
\end{verbatim}
        \item For Java 1.1, after downloading, type
\begin{verbatim}
jre -cp install_protege_4.0.2.jar install
\end{verbatim}
        \item If that does not work, try
\begin{verbatim}
java -classpath [path to] classes.zip:install_protege_4.0.2.jar install
\end{verbatim}
        \item If that does not work either, on sh-like shells, try
\begin{verbatim}
cd [to directory where install_protege_4.0.2.jar is located]
CLASSPATH=install_protege_4.0.2.jar
export CLASSPATH
java install
\end{verbatim}
        \item Or for csh-like shells, try
\begin{verbatim}
cd [to directory where install_protege_4.0.2.jar is located]
setenv CLASSPATH install_protege_4.0.2.jar
java install
\end{verbatim}
    \end{itemize}
    \item Instructions for other platforms
    \begin{itemize}
        \setlength{\itemsep}{0cm}
        \setlength{\parskip}{0cm}

        \item Be sure you have Java installed. You can download Java from Sun's website \cite{SunHome}.
        \item In a console window, change to the directory where you downloaded \texttt{install\_protege\_4.0.2.jar} to before running the installer.
    \end{itemize}
\end{enumerate}

Your operating system may invoke Java in a different way. To start the installer, add \texttt{install\_protege\_4.0.2.jar} to your \texttt{CLASSPATH}, then start the main class of the installer named \texttt{install}.

\newpage
\subsection{Plugins installation}
\label{sub:pluginsInstallation}

\begin{figure}[htp]
\centering
\includegraphics[scale=0.7]{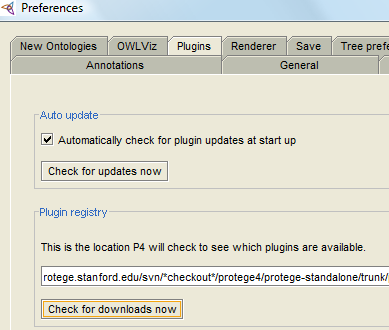}
\caption{Plugins tab}
\label{fig:plugins}
\end{figure}

After installation of Protégé there can be custom need of installation additional plugins. For plugins installation go to \textit{File->Preferences->Plugins} tab and click \textit{Check for downloads now} button (Figure \ref{fig:plugins}). 

\medskip

For all additional instructions check the wiki page for Protégé \cite{ProtegeWiki}. That is very competitive and reliable source of all information that may be needed. It includes documentation, tutorials, sample ontologies, plugin libraries, etc.

\section{Maven installation}
\label{sec:mavenInstallation}

Maven will be very helpful while working with sources of the project, because it can automatically download all dependencies tree from the network.

\bigskip

Maven is a Java tool, so you must have Java installed in order to proceed. Java Development Kit (JDK) is required (Java Runtime Environment (JRE) is not sufficient). Installation instructions provided below can be found on Maven home page \cite{MavenHome}.

\newpage

{\scshape Windows}

\begin{enumerate}
    \setlength{\itemsep}{0cm}
    \setlength{\parskip}{0cm}

    \item Download the latest Maven archive from Maven home page and unzip the distribution archive, i.e. \texttt{apache-maven-2.2.1-bin.zip} to the directory you wish to install Maven 2.2.1. These instructions assume you chose \path{C:\Program Files\Apache Software Foundation\}.
    \item Add the \texttt{M2\_HOME} environment variable by opening up the system properties (\texttt{WinKey+Pause}), selecting the \textit{Advanced} tab, and the \textit{Environment Variables} button, then adding the \texttt{M2\_HOME} variable in the user variables with the value \path{C:\Program Files\Apache Software Foundation\apache-maven-2.2.1}.
    \item In the same dialog, add the \texttt{M2} environment variable in the user variables with the value \path{%M2_HOME%\bin}.
    \item In the same dialog, update/create the \texttt{PATH} environment variable in the user variables and prepend the value \texttt{\%M2\%} to it (in order to make Maven available in the command line).
    \item In the same dialog, make sure that \texttt{JAVA\_HOME} exists in your user variables or in the system variables and it is set to the location of your JDK, e.g. \path{C:\Program Files\Java\jdk1.5.0_02} and that \path{%JAVA_HOME%\bin} is in your \texttt{PATH} environment variable.
\end{enumerate}

{\scshape Unix-based operating systems}

\begin{enumerate}
    \setlength{\itemsep}{0cm}
    \setlength{\parskip}{0cm}

    \item Download the latest Maven archive from Maven home page and unzip the distribution archive, i.e. \texttt{apache-maven-2.2.1-bin.zip} to the directory you wish to install Maven 2.2.1. These instructions assume you chose \path{/usr/local/apache-maven/}. The subdirectory \texttt{apache-maven-2.2.1} will be created from the archive.
    \item In a command terminal, add the \texttt{M2\_HOME} environment variable, e.g.
\begin{verbatim}
export M2_HOME=/usr/local/apache-maven/apache-maven-2.2.1
\end{verbatim}
    \item Add the \texttt{M2} environment variable, e.g. 
\begin{verbatim}
export M2=$M2_HOME/bin
\end{verbatim}
    \item Add \texttt{M2} environment variable to your path, e.g. 
\begin{verbatim}
export PATH=$M2:$PATH
\end{verbatim}
    \item Make sure that \texttt{JAVA\_HOME} is set to the location of your JDK, e.g. 
\begin{verbatim}
export JAVA_HOME=/usr/java/jdk1.5.0_02
\end{verbatim}
and that \path{$JAVA_HOME/bin} is in your \texttt{PATH} environment variable.   
\end{enumerate}

\setlength{\parindent}{0in}	

\chapter{Working with sources}
\label{cha:workingWithSources}

\indent When it comes to editing sources (for reasons like extending current functionality, fixing unexpected bug, etc.), we should prepare convenient developing environment. Project is written using Eclipse IDE, and is supported by Maven (installation steps in Section \ref{sec:mavenInstallation}). For Eclipse to support Maven, we should install M2Eclipse plugin. Installation is straightforward, because it is automatic, and can be found on M2Eclipse home page \cite{M2EclipseHome}. For me, that 3 mentioned tools are synonym of convenient developing environment in the Java world.

\bigskip

Traffic danger application consists of 3 projects:
\begin{itemize}
    \setlength{\itemsep}{0cm}
    \setlength{\parskip}{0cm}

    \item \texttt{traffic\_ont} - contains logic responsible for interacting with ontologies,
    \item \texttt{traffic\_domain} - contains logic responsible for interacting with database,
    \item \texttt{traffic\_web} - integration part, contains web logic, and UI.
\end{itemize}

\smallskip

For opening the sources, we should import that 3 projects into Eclipse workspace. When edition is complete, we can compile and install project to our local repository using Maven. What is more, complete web archive will be created, with all dependent resources and libraries. Such prepared package can be then deployed through Tomcat manager (as shown in Section \ref{sec:loadingAndConfiguration}).

\bigskip

Because of comprehensive configuration files and appropriate projects structure (compatible with Maven standard layout \cite{MavenStandardLayout}), installation can be accomplished in 3 steps shown below (suppose, that all projects are located in \path{C:\workspace} directory):

\begin{verbatim}
cd C:\workspace\traffic_ont\
mvn clean install
cd C:\workspace\traffic_domain\
mvn clean install
cd C:\workspace\traffic_web\
mvn clean install
\end{verbatim}

\newpage

After execution of that trivial commands, in our local Maven repository, suppose it is \path{C:\.m2\repository\}, a subdirectory  containing 3 projects will be created. Directory structure looks like that:

\begin{verbatim}
|--repository
   |--...
   |--agh
      |--traffic
         |--traffic_ont
            |--1.0.0
               |--traffic_ont-1.0.0.jar
               |--...
         |--traffic_domain
            |--1.0.0
               |--traffic_domain-1.0.0.jar
               |--...
         |--traffic_web
            |--1.0.0
               |--traffic_web-1.0.0.war
               |--...
\end{verbatim}

The complete web archive \texttt{traffic\_web-1.0.0.war} is now created, and ready to be deployed on server (see Appendix \ref{cha:systemDeployment}).

\bigskip

Alternative way of quick loading Maven webapp project into Tomcat servlet container is available by invoking following command:

\begin{verbatim}
mvn tomcat:run
\end{verbatim}

\clearpage
\listoffigures
\clearpage 
\listoftables

\bibliography{bibliography}

\begin{thebibliography}{10}

\bibitem{HermiTHome}
{HermiT} reasoner home page.
\newblock \url{http://hermit-reasoner.com/}.
\newblock Accessed September 2010.

\bibitem{HibernateDocs}
{Hibernate} documentation.
\newblock \url{http://docs.jboss.org/hibernate/stable/core/reference/en/html/}.
\newblock Accessed September 2010.

\bibitem{HibernateHome}
{Hibernate} home page.
\newblock \url{http://www.hibernate.org/}.
\newblock Accessed September 2010.

\bibitem{SemWebIntro}
{Introduction} to {Ontologies} and {Semantic} {Web}.
\newblock \url{http://obitko.com/tutorials/ontologies-semantic-web/}.
\newblock Accessed September 2010.

\bibitem{MavenStandardLayout}
{Introduction} to the {Standard} {Directory} {Layout} for {Maven}.
\newblock
  \url{http://maven.apache.org/guides/introduction/introduction-to-the-standard-directory-layout.html}.
\newblock Accessed September 2010.

\bibitem{JSPHome}
{JavaServer} {Pages} home page.
\newblock \url{http://java.sun.com/products/jsp/}.
\newblock Accessed September 2010.

\bibitem{jQueryHome}
{jQuery} home page.
\newblock \url{http://jquery.com/}.
\newblock Accessed September 2010.

\bibitem{jQueryUsageStatistics}
{jQuery} usage statistics.
\newblock \url{http://trends.builtwith.com/javascript/jquery/}.
\newblock Accessed September 2010.

\bibitem{M2EclipseHome}
{M2Eclipse} home page.
\newblock \url{http://m2eclipse.sonatype.org/}.
\newblock Accessed September 2010.

\bibitem{AgileManifesto}
{Manifesto} for {Agile} {Software} {Development}.
\newblock \url{http://agilemanifesto.org/}.
\newblock Accessed September 2010.

\bibitem{MavenHome}
{Maven} home page.
\newblock \url{http://maven.apache.org/}.
\newblock Accessed September 2010.

\bibitem{OWLAPIHome}
{OWL} {API} home page.
\newblock \url{http://owlapi.sourceforge.net/}.
\newblock Accessed September 2010.

\bibitem{PostgresDocs}
{PostgreSQL} documentation.
\newblock \url{http://www.postgresql.org/docs/8.4/static/index.html}.
\newblock Accessed September 2010.

\bibitem{PostgresHome}
{PostgreSQL} home page.
\newblock \url{http://www.postgresql.org/}.
\newblock Accessed September 2010.

\bibitem{ProtegeHome}
{Protégé} home page.
\newblock \url{http://protege.stanford.edu/}.
\newblock Accessed September 2010.

\bibitem{ProtegeWiki}
{Protégé} wiki page.
\newblock \url{http://protegewiki.stanford.edu/}.
\newblock Accessed September 2010.

\bibitem{SemWebTutorial}
{Semantic} {Web} tutorial.
\newblock \url{http://www.w3schools.com/semweb/default.asp/}.
\newblock Accessed September 2010.

\bibitem{SpringHome}
{Spring} home page.
\newblock \url{http://www.springsource.org/}.
\newblock Accessed September 2010.

\bibitem{SpringMVCSunBlueprints}
{SpringMVC} {Java} {BluePrints}.
\newblock \url{http://java.sun.com/blueprints/patterns/MVC.html}.
\newblock Accessed September 2010.

\bibitem{SunHome}
{Sun} home page.
\newblock \url{http://java.sun.com/}.
\newblock Accessed September 2010.

\bibitem{CasestudioHome}
{Toad} {Data} {Modeler}.
\newblock \url{http://www.casestudio.com/enu/default.aspx}.
\newblock Accessed September 2010.

\bibitem{UsageOfJavaScriptForWebsites}
{Usage} of {JavaScript} libraries for websites.
\newblock
  \url{http://w3techs.com/technologies/overview/javascript_library/all/}.
\newblock Accessed September 2010.

\bibitem{SemWebActivity}
{W3C} {Semantic} {Web} {Activity}.
\newblock \url{http://www.w3.org/2001/sw/}.
\newblock Accessed September 2010.

\bibitem{KSL}
{Why} {Develop} an {Ontology}?
\newblock
  \url{http://www-ksl-svc.stanford.edu:5915/doc/frame-editor/guided-tour/why-develop-an-ontology.html}.
\newblock Accessed September 2010.

\bibitem{W3SchoolsXML}
{XML} tutorial.
\newblock \url{http://www.w3schools.com/xml/}.
\newblock Accessed September 2010.

\bibitem{Arv08}
F.~Arvidsson and A.~Flycht-Eriksson.
\newblock {O}ntologies {I}, November 2008.

\bibitem{BCM03}
Franz Baader, Diego Calvanese, Deborah McGuinness, Daniele Nardi, and Peter
  Patel-Schneider.
\newblock {\em {The} {Description} {Logic} {Handbook}: {Theory},
  implementation, and applications}.
\newblock Cambridge University Press, 2003.

\bibitem{BLF99}
Tim Berners-Lee and Mark Fischetti.
\newblock {\em {Weaving} the {Web}: {The} {Past}, {Present} and {Future} of the
  {World} {Wide} {Web} by its {Inventor}}.
\newblock Britain: Orion Business, 1999.

\bibitem{BHL01}
Tim Berners-Lee, James Hendler, and Ora Lassila.
\newblock {The} {Semantic} {Web}.
\newblock {\em Scientific American}, 2001.

\bibitem{W3CXML}
Tim Bray, Jean Paoli, Eve Maler, François Yergeau, and C.~M. Sperberg-McQueen.
\newblock {Extensible} {Markup} {Language} ({XML}) 1.0 ({Fifth} {Edition}).
\newblock {W3C Recommendation}, W3C, November 2008.
\newblock \url{http://www.w3.org/TR/2008/REC-xml-20081126/}.

\bibitem{FoxGru}
Mark~S. Fox and Michael Gruninger.
\newblock {\em {Enterprise} {Modeling}}.
\newblock The American Association for Artificial Intelligence, 445 Burgess
  Drive Menlo Park, California 94025.

\bibitem{PWGB}
Pieter~W. Groenendijk.
\newblock {Wordpress} blog.
\newblock \url{http://pietergroenendijk.wordpress.com/}.
\newblock Accessed September 2010.

\bibitem{Gru93}
Thomas~R. Gruber.
\newblock {A} {Translation} {Approach} to {Portable} {Ontology}
  {Specifications}.
\newblock {\em Knowledge Acquisition}, 5(2):199--220, June 1993.

\bibitem{RDFSchema}
Ramanathan~V. Guha and Dan Brickley.
\newblock {RDF} {Vocabulary} {Description} {Language} 1.0: {RDF} {Schema}.
\newblock {W3C} {Recommendation}, W3C, February 2004.
\newblock \url{http://www.w3.org/TR/2004/REC-rdf-schema-20040210/}.

\bibitem{OWLGuide}
Matthew Horridge.
\newblock {\em {A} {Practical} {Guide} {To} {Building} {OWL} {Ontologies}
  {Using} {Protégé} 4 and {CO-ODE} {Tools}}.
\newblock The University Of Manchester, 1.2 edition, March 2009.

\bibitem{Hor09}
Matthew Horridge and Sean Bechhofer.
\newblock {The} {OWL} {API}: {A} {Java} {API} for {Working} with {OWL} 2
  {Ontologies}.
\newblock Technical report, The University of Manchester, UK, October 2009.
\newblock 6th OWL Experienced and Directions Workshop, Chantilly, Virginia.

\bibitem{HorSSF}
Ian~R. Horrocks.
\newblock {Semantic} {Web}: {The} {Story} {So} {Far}.
\newblock University of Manchester, Manchester, UK.

\bibitem{Hor97}
Ian~R. Horrocks.
\newblock {\em {Optimising} {Tableaux} {Decision} {Procedures} {For}
  {Description Logics}}.
\newblock PhD thesis, University of Manchester, 1997.

\bibitem{UCEReport}
Aaron Kershenbaum, Achille Fokoue, Chintan Patel, Christopher Welty, Edith
  Schonberg, James Cimino, Li~Ma, Kavitha Srinivas, Robert Schloss, and
  J~William Murdock.
\newblock {A} {View} of {OWL} from the {Field}: {Use} cases and {Experiences}.
\newblock Technical report, IBM T. J. Watson Research Lab, Hawthorne, USA and
  IBM China Research Lab, Beijing, China and Department of Biomedical
  Informatics, Columbia University, New York, USA, 2006.

\bibitem{OntDrivDev}
Holger Knublauch.
\newblock {\em {Ontology-Driven} {Software} {Development} in the {Context} of
  the {Semantic} {Web}: {An} {Example} {Scenario} with {Protégé}/{OWL}}.
\newblock Stanford Medical Informatics, Stanford University, Stanford, CA,
  94305.

\bibitem{Mar08}
Robert~C. Martin and Micah Martin.
\newblock {\em {Agile} {Principles}, {Patterns}, and {Practices} in {C\#}}.
\newblock Helion, 2008.

\bibitem{RDFPrimer}
Eric Miller and Frank Manola.
\newblock {RDF} {Primer}.
\newblock {W3C} {Recommendation}, W3C, February 2004.
\newblock \url{http://www.w3.org/TR/2004/REC-rdf-primer-20040210/}.

\bibitem{GJNSemWeb}
Grzegorz~J. Nalepa.
\newblock {Semantic} {Web}.
\newblock Presentation at Institute of Automatics, AGH University of Science
  and Technology, Poland.

\bibitem{OntDev101}
Natalya~F. Noy and Deborah~L. McGuinness.
\newblock {\em {Ontology} {Development} 101: {A} {Guide} to {Creating} {Your}
  {First} {Ontology}}.
\newblock Stanford University, Stanford, CA, 94305.

\bibitem{Ree78}
Trygve Reenskaug.
\newblock {MVC} {XEROX} {PARC} 1978-79.

\bibitem{CLR90}
Ronald L.~Rivest Thomas H.~Cormen, Charles E.~Leiserson.
\newblock {\em {Introduction} to {Algorithms}}, volume~1.
\newblock The MIT Press and McGraw-Hill, 1990.

\bibitem{HLP08}
Frank van Harmelen, Vladimir Lifschitz, and Bruce Porter.
\newblock {\em {Handbook} of {Knowledge} {Representation} ({Foundations} of
  {Artificial} {Intelligence})}.
\newblock Elsevier B.V., 1 edition, 2008.

\bibitem{W3COWL}
{W3C OWL Working Group}.
\newblock {OWL} 2 {Web} {Ontology} {Language} {Document} {Overview}.
\newblock Technical report, W3C, October 2009.
\newblock \url{http://www.w3.org/TR/2009/REC-owl2-overview-20091027/}.

\bibitem{Wei76}
Karl Weick.
\newblock {Educational} organizations as loosely coupled systems.
\newblock {\em Administrative Science Quarterly}, 21:1--9, 1976.

\end{thebibliography}

\end{document}